\documentclass[
 amsmath,amssymb,
 aps,
prc,
]{revtex4-2}

\usepackage{graphicx}
\usepackage{dcolumn}
\usepackage{bm}
\usepackage{hyperref}

\usepackage{xcolor}

\usepackage[normalem]{ulem}

\begin{document}

\title{Quark flavor equilibration of the quark-gluon plasma}

\author{Andrew Gordeev\textsuperscript{1}, Steffen A. Bass\textsuperscript{1}, Berndt Müller\textsuperscript{1}, and Jean-François Paquet\textsuperscript{2}}
\affiliation{
 \textsuperscript{1}Department of Physics, Duke University, Durham, NC 27708, USA
 \\
 \textsuperscript{2}Department of Physics and Astronomy, Vanderbilt University, Nashville TN 37235, USA
}

\begin{abstract}
The early stage of a heavy-ion collision is marked by rapid entropy production and the transition from a gluon saturated initial condition to a plasma of quarks and gluons that evolves hydrodynamically. However, during the early times of the hydrodynamic evolution, the chemical composition of the QCD medium is still largely unknown. We present a study of quark chemical equilibration in the (Q)GP using a novel model of viscous hydrodynamic evolution in partial chemical equilibrium. Motivated by the success of gluon saturated initial condition models, we initialize the QCD medium as a completely gluon dominated state. Local quark production during the hydrodynamic phase is then simulated through the evolution of time-dependent fugacities for each independent quark flavor, with the timescales set as free parameters to compare different rates of equilibration. We present the results of complete heavy-ion collision simulations using this partial chemical equilibrium model, and show the effects on hadronic and electromagnetic observables. In particular, we show that the development of flow is sensitive to the equilibration timescale, providing an empirical way to probe the chemical equilibration of the QCD medium.
\end{abstract}

\maketitle
                              
\section{Introduction}

The initial impact of a relativistic heavy-ion collision generates a highly energetic medium far from thermodynamic equilibrium. It has long been an open question precisely how this early medium transitions to the quark-gluon plasma (QGP), a system well-described by hydrodynamics and thus near local thermodynamic equilibrium. The phenomenological success of theoretical models for the early time medium, such as IP-Glasma \cite{Schenke:2012wb} and EKRT \cite{Eskola:1999fc}, supports the concept of a gluon saturated initial state. In such a scenario, it is hypothesized that gluons are produced copiously to the point that gluon splitting and recombination processes reach a dynamic balance \cite{McLerran:1993ni, Gribov:1983ivg, Gelis:2010nm}.

There is compelling evidence that gluon thermalization occurs quite rapidly through ''bottom-up'' thermalization in the context of a weakly-coupled QGP \cite{Baier:2000sb}.
Early perturbative QCD studies of the equilibration timescales suggested that gluons should equilibrate more rapidly than quarks and antiquarks, due to significantly larger gluon-gluon cross sections as compared to quark-gluon or quark-quark cross sections \cite{Shuryak:1992wc,Biro:1993qt}.
Subsequent kinetic theory descriptions have tended to support rapid hydrodynamization on a timescale of $\approx$1 fm/c \cite{El:2007vg, Kurkela:2015qoa, Keegan:2016cpi, Kurkela:2018vqr}. Similar results have been found using classical-statistical real-time lattice simulations \cite{Epelbaum:2013ekf, Berges:2013fga, Schenke:2015aqa}. Two-stage equilibration models have shown that quark-antiquark chemical equilibration may follow on a timescale of over 3 fm/c, although still before local thermalization of the QGP \cite{Xu:2004mz, Kurkela:2018oqw, Kurkela:2018xxd}. Despite considerable uncertainty surrounding the precise timescale of quark-antiquark chemical equilibrium, these studies show that the QGP may form well before achieving chemical equilibration, potentially allowing for significant effects on its properties and evolution.

Previous studies have examined the effects of hydrodynamic evolution in partial chemical equilibrium \cite{Vovchenko:2015yia, Vovchenko:2016ijt}. In this regime, the medium forms in a gluon-saturated initial state and subsequently equilibrates according to a time-dependent equation of state (EoS). These studies have found sensitivity to chemical equilibration in photonic and dileptonic observables, suggesting that electromagnetic probes can reveal information about the early stages of the collision. However, to date, no prior work has extended this scheme to a precision study of hadronic observables, which may provide additional signals of the composition of the QGP and are extensively measured in experiments at the Large Hadron Collider (LHC) and Relativistic Heavy Ion Collider (RHIC). This paper aims to address this gap, presenting the results for simulations of complete heavy-ion collision events with a viscous QGP evolved out of chemical equilibrium.

The remainder of this paper is organized as follows: Section II describes the partial chemical equilibrium model for the QGP and its implementation in a heavy-ion collision event generator. Section III presents results from generated events using this model, showing the effects on both hadronic and electromagnetic observables. Section IV summarizes our findings and outlines directions for future research.

\section{Modeling Partial Chemical Equilibrium}

Our model hydrodynamically evolves the QCD medium in partial chemical equilibrium; i.e., with fully thermalized gluons but zero initial (anti)quark content. This requires a particular construction of the equation of state, but does not directly modify the fundamental equations of motion of relativistic hydrodynamics:

\begin{align}
    \partial_\mu T^{\mu \nu} = 0,
\end{align}

where we define the energy-momentum tensor $T^{\mu \nu}$ as

\begin{align}
    T^{\mu \nu} = \varepsilon u^\mu u^\nu - (P + \Pi) (g^{\mu \nu} - u^\mu u^\nu) + \pi^{\mu \nu},
\end{align}

with $\varepsilon$ denoting the energy density, $u^\mu$ the local flow velocity, $P$ the pressure, $\Pi$ the bulk pressure, $g^{\mu \nu}$ the metric tensor (using the mostly-minus convention $g^{\mu \nu} = \text{diag}(+1,-1,-1,-1)$), and $\pi^{\mu \nu}$ the shear stress tensor. In this work, we use Israel-Stewart-type second-order viscous hydrodynamics, with additional equations of motion for the evolution of $\Pi$ and $\pi^{\mu \nu}$ detailed in \cite{Denicol:2012cn, Molnar:2013lta, Denicol:2014vaa}. With the typical assumption of chemical equilibrium, $P$ and $\varepsilon$ are related by a fixed equation of state, while $\Pi$ and $\pi^{\mu \nu}$ reflect deviations from local thermodynamic equilibrium. However, in partial chemical equilibrium, all of these quantities can change with the local quark fugacity.

\subsection{Equation of State}

The equation of state encodes the chemistry of the QGP by relating the primary state variables that define the medium. Thus, defining the non-equilibrium EoS is essential to evolving the medium out of chemical equilibrium. We do so by calculating the pressure $P$ and energy density $\varepsilon$ as functions of the temperature $T$ and a quark fugacity $\gamma_q$. This is sufficient to define the EoS, although we additionally calculate the entropy density $s$ in Appendix \ref{sec:s}.

\subsubsection{High Temperature}

In equilibrium, the EoS at high temperatures is derived from lattice QCD calculations. Ideally, we would hope to calculate the non-equilibrium EoS the same way. However, due to the numerical sign problem, such a calculation is thus far impractical, and lattice studies have largely been limited to equilibrium calculations at finite but small chemical potentials \cite{Bazavov:2017dus}. Instead, we adopt an approach based on that proposed in \cite{Vovchenko:2016ijt} which interpolates between two equilibrium lattice calculations: one in (2+1)-flavor QCD \cite{HotQCD:2014kol}, and one in pure SU(3) gauge theory \cite{Borsanyi:2012ve}. 
One should note that that pure SU(3) gauge theory is only an approximation for a gluon-dominated QCD system. Even in an out-of-chemical-equilibrium environment where gluons predominate and quarks are scarce, virtual quark contributions still impact gluon interactions and the running of the strong coupling constant.

We assume that quarks and antiquarks have equal densities characterized by the quark fugacity $\gamma_q$, which linearly interpolates the pressure and energy density of the two lattice equations of state in the following way:

\begin{equation}
\begin{aligned}
\frac{P}{T^4}(T,\gamma_q) = \gamma_q \frac{P_3}{T^4}
   \left(T \frac{T_3}{T_{\mathrm{c}}(\gamma_q)}\right) + (1-\gamma_q) \frac{P_0}{T^4}
   \left(T \frac{T_0}{T_{\mathrm{c}}(\gamma_q)}\right),
\end{aligned}
\end{equation}

\begin{equation}
\begin{aligned}
\frac{\varepsilon}{T^4}(T,\gamma_q) = \gamma_q \frac{\varepsilon_3}{T^4}
   \left(T \frac{T_3}{T_{\mathrm{c}}(\gamma_q)}\right) + (1-\gamma_q) \frac{\varepsilon_0}{T^4}
   \left(T \frac{T_0}{T_{\mathrm{c}}(\gamma_q)}\right) .
\end{aligned}
\end{equation}

where $P_3 (T)$, $\varepsilon_3 (T)$, and $T_3$ are respectively the pressure, energy density, and critical temperature of the full QCD EoS, and $P_0 (T)$, $\varepsilon_0 (T)$, and $T_0$ are the corresponding quantities for the pure glue EoS. Note that at $\gamma_q = 1$, we recover exactly the (2+1)-flavor QCD EoS, and at $\gamma_q = 0$, we recover exactly the gluonic EoS. This quark fugacity can be easily generalized to flavor-specific fugacities, but in this work we will consider only the simplified case of a single $\gamma_q$ that encompasses all quark flavors.

As there is known to be a first-order phase transition in $N_f = 0$ at $T_0 \approx 260$ MeV, as opposed to a crossover in $N_f = 2+1$ at $T_3 \approx 158$ MeV, simply interpolating between $\frac{P_3}{T^4} (T)$ and $\frac{P_0}{T^4} (T)$ without additional modifications would cause the system to experience separate transitions at each critical temperature for any $0 < \gamma_q < 1$. To avoid this, we have rescaled all temperatures according to a fugacity-dependent critical temperature $T_\mathrm{c}(\gamma_q)$ so that there is only ever one transition at a given fugacity \cite{Moreau:unpublished}. $T_\mathrm{c}(\gamma_q)$ is defined as 

\begin{align}
    T_\mathrm{c}(\gamma_q) = \sqrt{\gamma_q} T_3 + (1-\sqrt{\gamma_q}) T_0.
    \label{eq:Tc}
\end{align}

\begin{figure}[!htbp]
    \centering
    \includegraphics[width=0.6\linewidth]{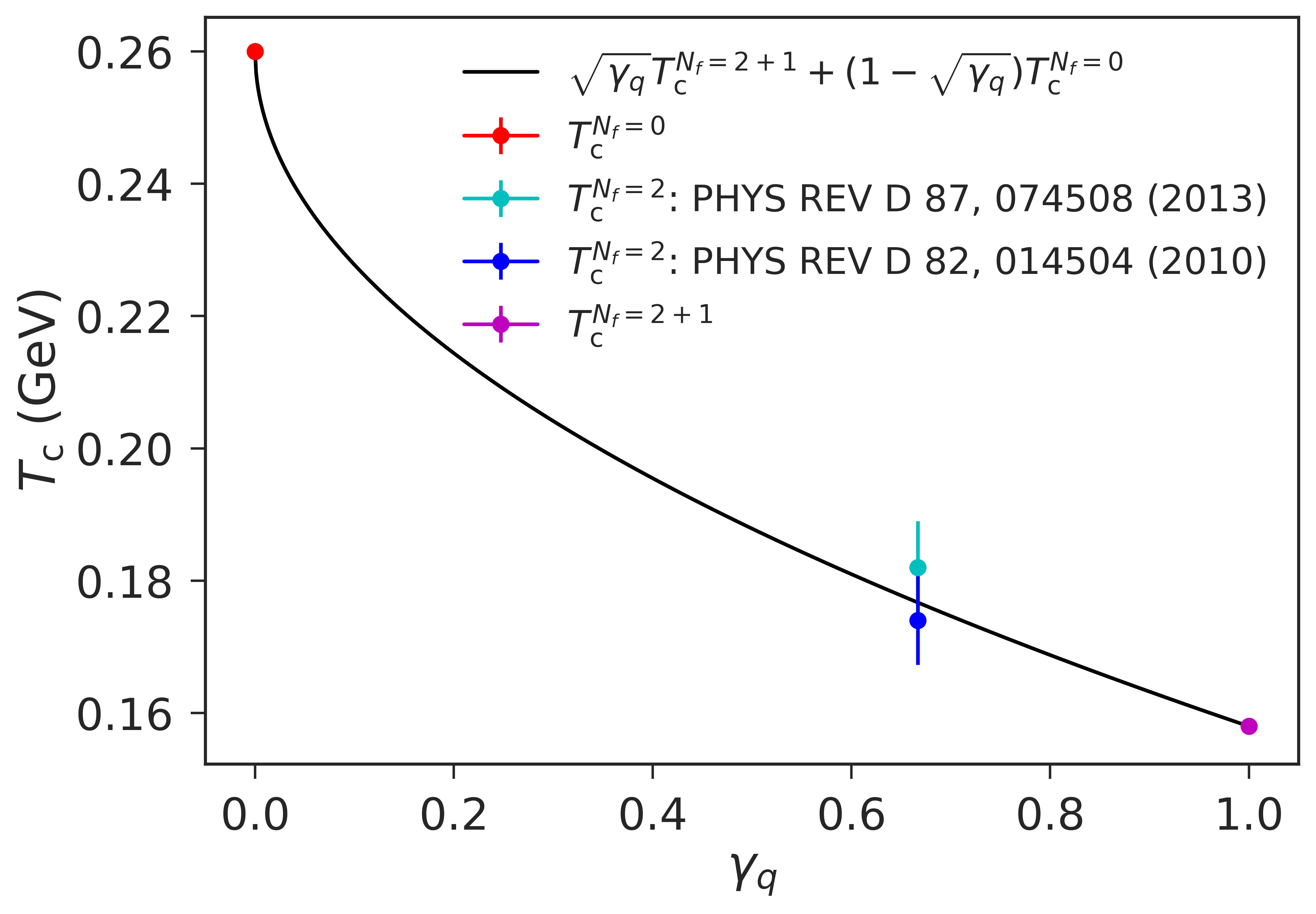}
    \caption{$T_\mathrm{c}(T,\gamma_q)$ compared to the critical temperatures for $N_f = 0$ \cite{Borsanyi:2012ve}, $N_f = 2$ \cite{Burger:2011zc, Bornyakov:2009qh}, and $N_f = 2+1$ \cite{HotQCD:2014kol} results.}
    \label{fig:Tc}
\end{figure}

The behavior of the phase transition at intermediate quark fugacities is largely unknown. As a guiding approximation, we construct the fugacity-dependent critical temperature such that it smoothly interpolates between the lattice QCD results for $T_0$ and $T_3$ while passing through the $N_f = 2$ critical temperature \cite{Burger:2011zc, Bornyakov:2009qh} at $\gamma_q = 2/3$, although it should be noted that these $N_f = 2$ results are derived with heavier-than-physical quark masses. Fig. \ref{fig:Tc} plots the result. 

Note that with this temperature rescaling, the actual pressure $P (T,\gamma_q)$ and energy density $\varepsilon (T, \gamma_q)$ take the forms

\begin{equation}
\begin{aligned}
P(T,\gamma_q) = \gamma_q
   \left(\frac{T_{\mathrm{c}}(\gamma_q)}{T_3}\right)^4
   P_3\!\left(T \frac{T_3}{T_{\mathrm{c}}(\gamma_q)}\right) + (1-\gamma_q)
   \left(\frac{T_{\mathrm{c}}(\gamma_q)}{T_0}\right)^4
   P_0\!\left(T \frac{T_0}{T_{\mathrm{c}}(\gamma_q)}\right) ,
\end{aligned}
\label{eq:P}
\end{equation}

\begin{equation}
\begin{aligned}
\varepsilon(T,\gamma_q) = \gamma_q
   \left(\frac{T_{\mathrm{c}}(\gamma_q)}{T_3}\right)^4
   \varepsilon_3\!\left(T \frac{T_3}{T_{\mathrm{c}}(\gamma_q)}\right) + (1-\gamma_q)
   \left(\frac{T_{\mathrm{c}}(\gamma_q)}{T_0}\right)^4
   \varepsilon_0\!\left(T \frac{T_0}{T_{\mathrm{c}}(\gamma_q)}\right) .
\end{aligned}
\end{equation}

\subsubsection{Low Temperature}

Conventionally, the lattice equation of state at high temperature is matched to a hadron resonance gas equation of state at low temperature \cite{Huovinen:2009yb}. We do the same here, only with modifications to the hadron distribution functions to account for non-zero quark chemical potential. In principle, the low temperature pure glue EoS corresponds to gas of glueballs \cite{Stoecker:2015zea}, rather than a gas of hadrons. With the addition of a Hagedorn spectrum \cite{Meyer:2009tq}, the glueball gas model agrees well with lattice results while additionally explaining the very low pressure and energy density at $T < T_\mathrm{c}$ through the large masses of the constituent glueballs \cite{Borsanyi:2012ve}. However, in this work we use the hadron resonance gas equation of state for all $0 \leq \gamma_q \leq 1$, and instead introduce hadronic fugacities that suppress the hadron distributions at low $\gamma_q$.

The hadron resonance gas energy density and pressure are calculated as:
\begin{align}
    \varepsilon &= \sum_i g_i \int \frac{d^3 p}{(2\pi)^3} E_p f_i(p) \\
    P &= \sum_i g_i \int \frac{d^3 p}{(2\pi)^3} \frac{p^2}{3E_p} f_i(p),
    \label{eq:ep_hrg}
\end{align}
where $i$ denotes the hadron species, $g_i$ and $f_i$ are the respective degeneracy and distribution function for each species, and $E_p = \sqrt{p^2 + m_i^2}$ is the individual hadron energy given a mass of $m_i$. In equilibrium, mesons are described by a Bose-Einstein distribution and baryons by a Fermi-Dirac distribution. In chemical non-equilibrium, the distribution functions can be modified to take the form
\begin{align}
    f_i(p, \lambda_i) = \frac{1}{\lambda_i^{-1} e^{E_p/T} \pm 1},
    \label{eq:f_i}
\end{align}
where $\lambda_i$ is a species-specific fugacity factor distinct from the $\gamma_q$ used in the high temperature EoS.

There is some ambiguity in how the hadronic fugacities should map to the quark fugacities. One constraint is that as baryons have three (anti)quarks and mesons have two, we expect that baryons should be further suppressed by a power of 3/2. Another constraint is that we seek to construct an EoS where all thermodynamic variables are continuous and differentiable over the transition region, for all quark fugacities between 0 and 1. These constraints lead us to define the hadronic fugacities as

\begin{align}
    \lambda_{\text{meson}} &= 0.85 \gamma_q + 0.15 \\
    \lambda_{\text{baryon}} &= \lambda_{\text{meson}}^{3/2},
    \label{eq:lambda_i}
\end{align}

where the numerical constants are a fit to ensure smooth matching with the high temperature EoS. Using these hadronic fugacities, one can use the distribution functions in Eq. \ref{eq:f_i} in Eq. \ref{eq:ep_hrg} to construct the lower temperature EoS.

\begin{figure*}[htbp]
    \centering
    \includegraphics[width=0.49\textwidth]{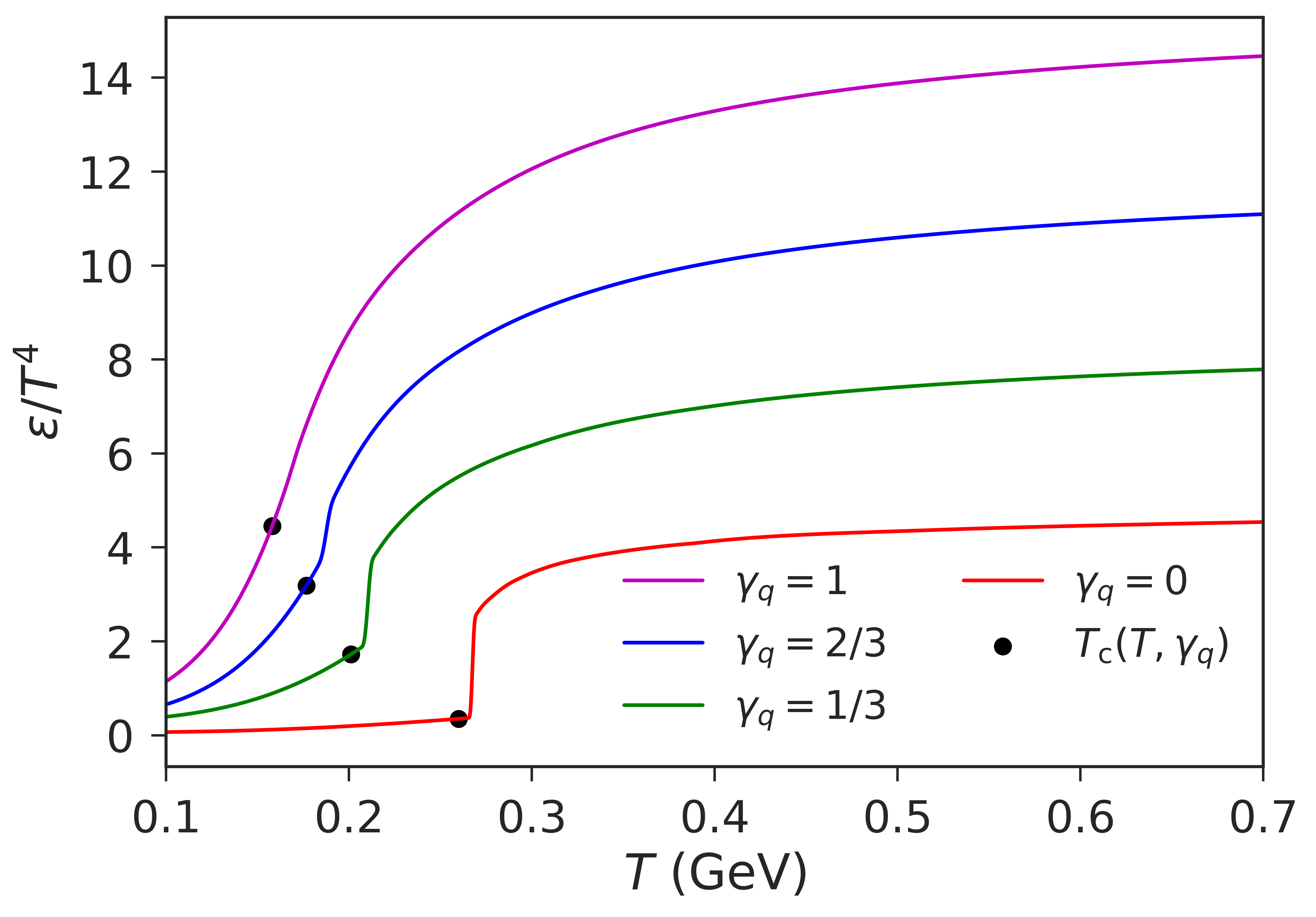}%
    \hfill
    \includegraphics[width=0.49\textwidth]{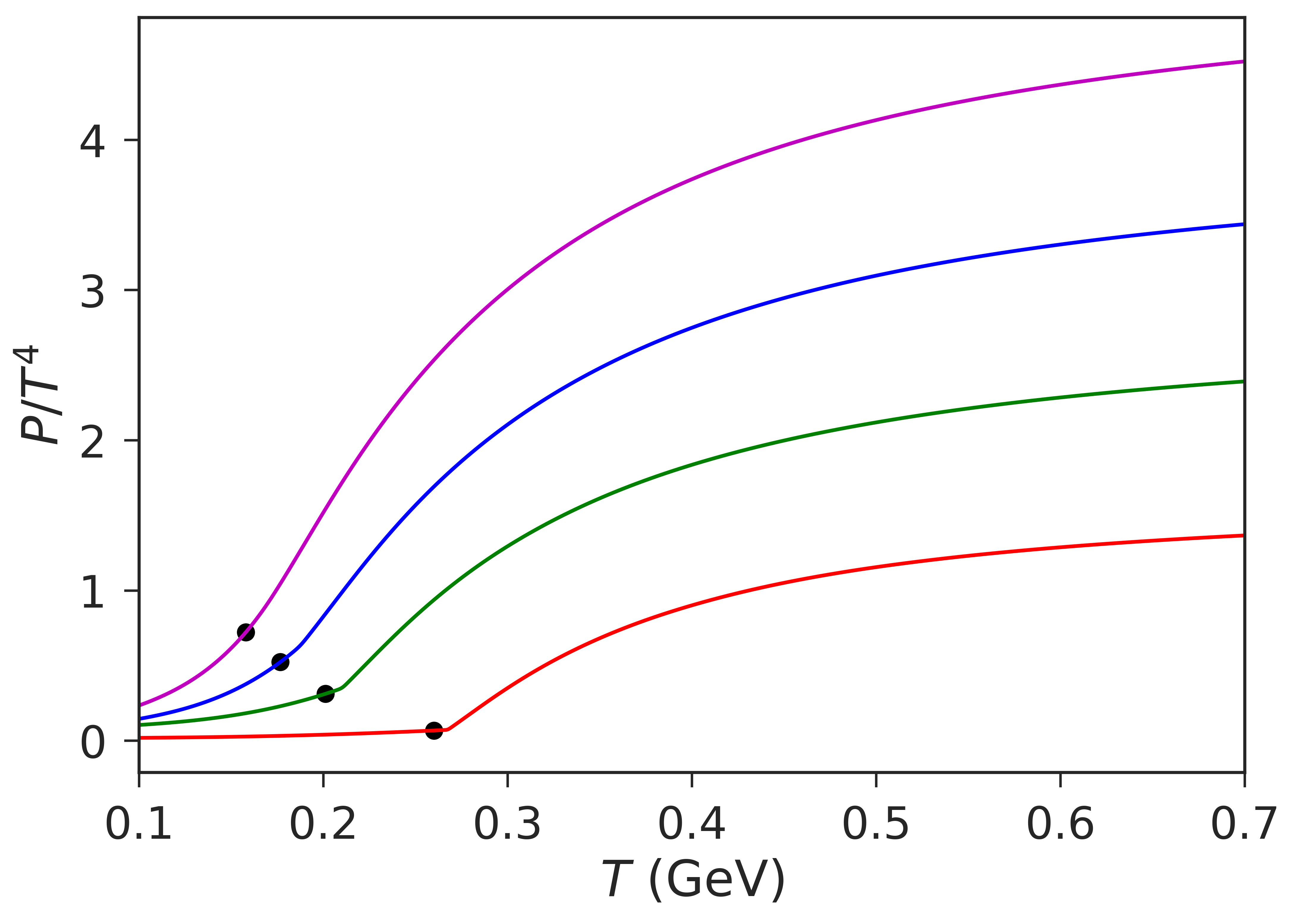}
    \caption{Energy density (left) and pressure (right) of the partial chemical equilibrium EoS, constructed by matching a linear interpolation in $\gamma_q$ of lattice data at high $T$ to a non-equilibrium hadron resonance gas low $T$. Note the first-order phase transition in the pure glue EoS at $T_\mathrm{c} = 260$ MeV. The black points indicate the respective critical temperature $T_\mathrm{c} (\gamma_q)$ for each value of $\gamma_q$.}
    \label{fig:EoS}
\end{figure*}

This low temperature result for $\varepsilon(T,\gamma_q)$ and $P(T,\gamma_q)$ is conventionally matched to the high temperature values over a crossover range of $T \approx 160-200 $ MeV. We do the same, using Krogh interpolation, only for a fugacity-dependent temperature range defined as $(T_\mathrm{c}(\gamma_q) + 5 $ MeV$, T_\mathrm{c}(\gamma_q) + (10 + 100 \gamma_q) $ MeV$).$ Fig. \ref{fig:EoS} shows the resulting $\varepsilon(T,\gamma_q)$ and $P(T,\gamma_q)$ for several choices of $\gamma_q$. 

Because we interpolate with the pure glue equation of state for all intermediate fugacities, there will be a first-order phase transition for all $\gamma_q < 1$ with a magnitude $(1-\gamma_q)$. We do not observe a significant impact of this transition on the QGP evolution, however, as we particlize at $T_\mathrm{c} (\gamma_q)$. This is discussed further in Sec. \ref{sec:particlization}.

\subsection{Quark Fugacity}

\begin{figure}[!htbp]
    \centering
    \includegraphics[width=0.7\linewidth]{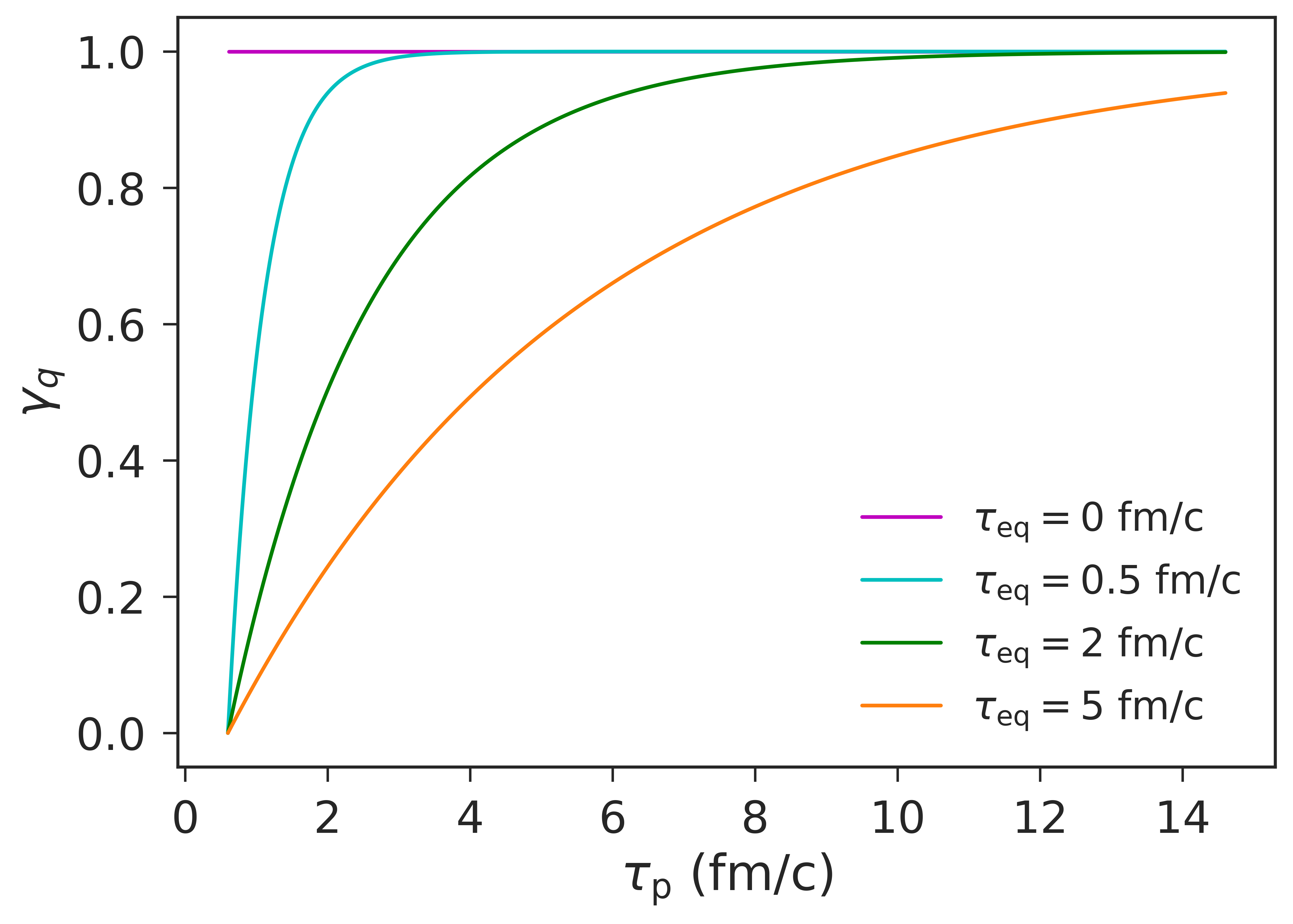}
    \caption{$\gamma_q$ as a function of proper time for several equilibration timescales $\tau_\mathrm{eq}$ for $\tau_0 = 0.6$ fm/c.}
    \label{fig:fugacity}
\end{figure}

As defined above, the equation of state in partial chemical equilibrium depends on two variables: the temperature $T$ and the quark fugacity $\gamma_q$. In practice, we invert $\varepsilon(T,\gamma_q)$ to treat $\varepsilon$ as an independent field which is evolved hydrodynamically. One should, in principle, introduce additional rate equations that govern the time evolution of the quark fugacity $\gamma_q$ \cite{Biro:1993qt}. To allow for more direct control of the equilibration timescale, we instead parameterize $\gamma_q$ as a simple function of the local proper time $\tau_\mathrm{p}$:

\begin{align}
    \gamma_q(\tau_\mathrm{p}) = 1 - \exp\left(\frac{\tau_0 - \tau_\mathrm{p}}{\tau_\mathrm{eq}}\right),
    \label{eq:fugacity}
\end{align}

where $\tau_0$ is the global initial time of the system and $\tau_\mathrm{eq}$ is a free parameter corresponding to the effective chemical equilibration time. Fig. \ref{fig:fugacity} shows the form of this function.

Note that $\tau_\mathrm{p}$ is distinct from the time coordinate $\tau$; due to transverse flow and time dilation, the local proper time of each fluid cell evolves at a different rate. Thus, during the hydrodynamic evolution we must solve the additional equation
\begin{align}
    u^\mu \partial_\mu \tau_\mathrm{p} = 1
    \label{eq:proper_time}
\end{align}
with the initial condition $\tau_\mathrm{p}(\tau_0) = \tau_0$ to correctly determine the fugacity, and thus EoS, everywhere in the fluid. We treat $\tau_\mathrm{p}$ as an auxiliary field and evolve it alongside $\varepsilon$ and $\gamma_q$. At each timestep, while $T^{\mu \nu}$ and its constituent fields are updated according to Israel-Stewart-type viscous hydrodynamics, $\tau_\mathrm{p}$ is updated by advecting it along the fluid velocity $u^\mu$. The result of this is that regions of the fluid with greater flow velocities will evolve in proper time more gradually. Consequently, the periphery of the system will typically chemically equilibrate more slowly than the center in the lab frame.

\subsection{Particlization}
\label{sec:particlization}

\begin{figure}[!htbp]
    \centering
    \includegraphics[width=0.7\linewidth]{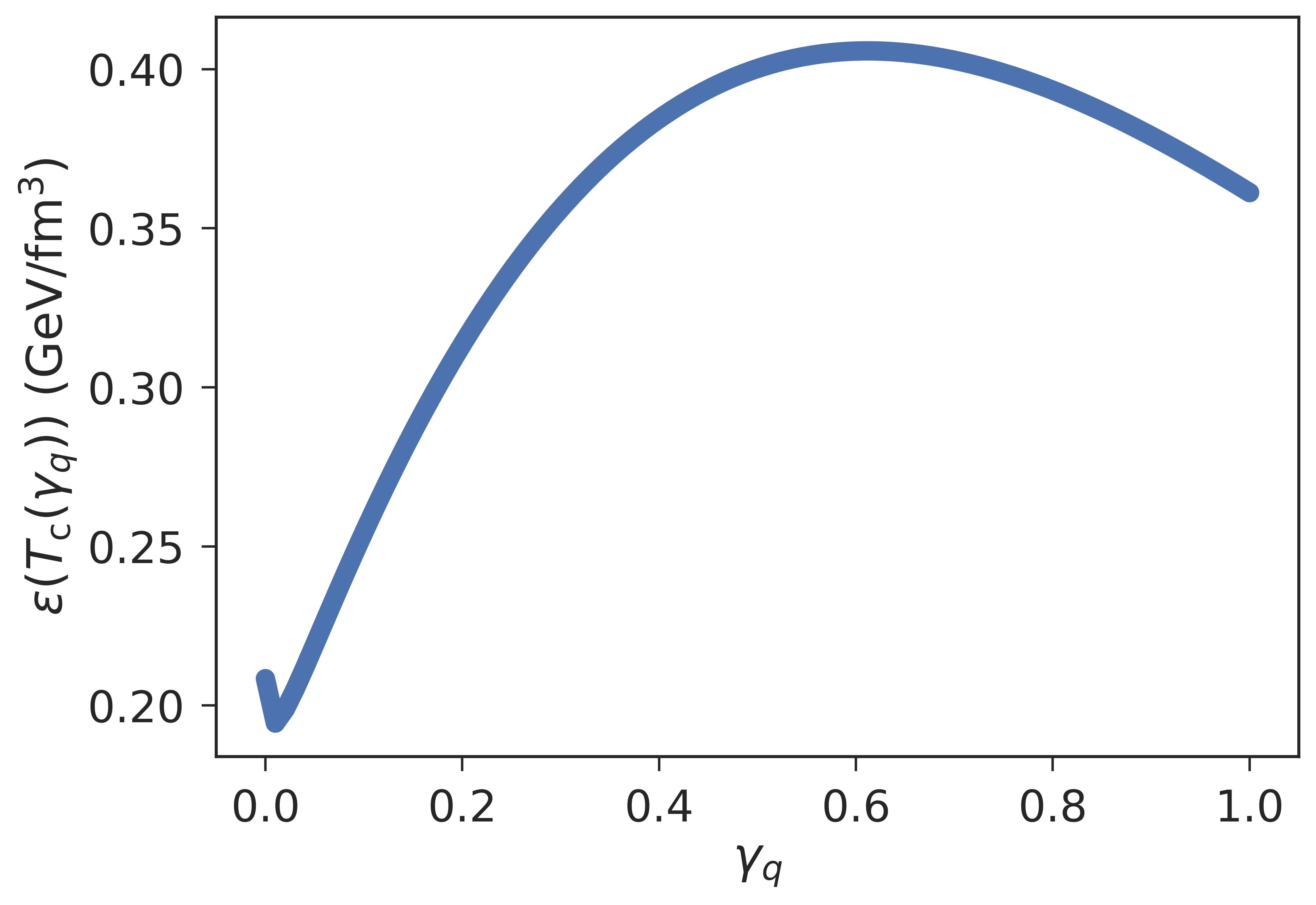}
    \caption{Energy density along the $T_\mathrm{c}(\gamma_q)$ hypersurface plotted against $\gamma_q$.}
    \label{fig:e_Tc}
\end{figure}

After hydrodynamics, we use the Cooper-Frye prescription \cite{Cooper:1974mv} to convert from a continuous fluid to a set of discrete particles, given by

\begin{align}
    \frac{d^3N_i}{dp^3} = g_i \int_\Sigma \frac{d\Sigma_\mu p^\mu}{(2\pi)^3} f_i (x,p), 
\end{align}

where $i$ denotes each hadron species, $g_i$ their respective degeneracies, and $\Sigma$ the particlization hypersurface with normal surface elements $d\Sigma_\mu$. The hypersurface $\Sigma$ is typically defined by a condition of constant temperature or energy density, and constructed from all of the fluid cells crossing that threshold \cite{Huovinen:2012is}. To account for the fugacity-dependent nature of the transition in our equation of state, we instead define $\Sigma$ according to the critical temperature $T_\mathrm{c}(\gamma_q)$, as given in Eq. \ref{eq:Tc}. The primary effect of this is that fluid cells with low fugacities will both hadronize and particlize at higher temperatures. Fig. \ref{fig:e_Tc} shows the corresponding energy densities at $T_\mathrm{c}(\gamma_q)$. $\varepsilon(T_\mathrm{c}(\gamma_q))$ changes non-monotonically with $\gamma_q$ as a result of our interpolation scheme, and can vary by as much as $\approx 50\%$. As shown in Sec. \ref{sec:evolution}, though, few fluid cells will be at fugacities near zero for all but the largest equilibration timescales, and the variation in practice is typically $\approx 20\%$. 

In ideal hydrodynamics, the only other significant modification that needs to be made is that the hadron distribution functions $f_i(x,p)$ must be modified with fugacity factors, as in Eq. \ref{eq:f_i}. This is equivalent to requiring that the equation of state remain consistent between the hydrodynamic and particlized regimes.

In viscous hydrodynamics, we additionally need to account for viscous corrections to the distribution functions, defining 

\begin{align}
    f_i(x,p) = f_{eq,i}(x,p) + \delta f_i(x,p),
\end{align}

where $f_{eq,i}$ are the equilibrium distribution functions and $\delta f_i$ are the corrections. Although the medium is in partial chemical equilibrium, we assume that it is close enough to local thermodynamic equilibrium for hydrodynamics to apply and identify $f_{eq,i}$ with the fugacity-modified distribution functions given by Eq. \ref{eq:f_i}. $\delta f_i$ is commonly linearized in the shear stress tensor $\pi^{\mu \nu}$, bulk pressure $\Pi$, and any conserved charge currents (typically the baryon diffusion current $V^\mu_B)$ \cite{Denicol:2012cn}. Note that quark number is explicitly not conserved. While we produce quarks and antiquarks equally, we exclusively work with $\mu_B = 0$ so that there is no nontrivial net baryon charge. Thus, any viscous correction must be accounted for in the shear and bulk correction terms. 

We choose the 14-moment approximation, where $\delta f_i$ is expanded in momentum moments of the distribution function and truncated after second-order terms in momentum \cite{Grad:1949zza, Denicol:2012cn, Denicol:2014vaa}. Using the notation $\Theta_i = 1$ ($\Theta_i = -1$) for bosons (fermions), we define

\begin{align}
    \bar{f}_{eq,i} = 1 - g_i^{-1} \Theta_i f_{eq,i}.
\end{align}

With zero net baryon current, the 14-moment expansion then takes the form

\begin{align}
    \delta f_i = f_{eq,i} \bar{f}_{eq,i} (c_T m_i^2 + c_E (u \cdot p)^2 + c_\pi^{\langle\mu\nu\rangle} p_{\langle\mu}p_{\nu\rangle}),
\end{align}

where $c_T$, $c_E$, and $c_\pi^{\langle\mu\nu\rangle}$ are the nonzero expansion coefficients \cite{Monnai:2009ad, McNelis:2021acu}. The notation $A^{\langle \mu_1 ... \mu_n \rangle}$ denotes the symmetric, traceless, and orthogonal projection of the tensor $A^{\mu_1 ... \mu_n}$ with respect to $u^\mu$. It can be shown that $c_T$, $c_E$, and $c_\pi^{\langle\mu\nu\rangle}$ are all calculable in terms of $\Pi$, $\pi^{\mu \nu}$, and thermal integrals of the form

\begin{align}
    J_{kq,i} = \int_p \frac{(u \cdot p)^{k-2q}(-p\cdot \Delta \cdot p)^q}{(2q+1)!!} f_{eq,i} \bar{f}_{eq,i}.
\end{align}

When particlizing a fluid cell with arbitrary fugacity $\gamma_q$, we use the 14-moment expansion coefficients corresponding to $J_{kq,i}$ calculated with the respective $f_{eq,i}(\gamma_q).$ In principle, one should expect the shear and bulk viscosities, and thus $\Pi$ and $\pi^{\mu \nu}$, to vary with the quark fugacity as well. However, due the difficulty of determining this effect, we neglect any such fugacity dependence here. 

\subsection{Implementation}

We simulate Pb+Pb collision events using a multi-stage model that incorporates Monte Carlo initial condition generation, hydrodynamic evolution, particlization, and hadronic transport.

We generate initial energy density profiles using T\textsubscript{R}ENTo \cite{Moreland:2014oya}, a parametric initial condition model . These profiles are initialized at $\tau_0 = 0.6$ fm/c with zero transverse flow. The T\textsubscript{R}ENTo parameters were set without specific calibration to experimental multiplicities at a particular center-of-mass energy, as the aim here is to demonstrate the effects of varying the chemical equilibration timescale rather than empirically constrain the model parameters. As a result, the simulated charged-particle multiplicities are intermediate between those measured at $\sqrt{s_{NN}} = 2.76$ TeV and $\sqrt{s_{NN}} = 5.02$ TeV.

To demonstrate the effects of quark chemical equilibration, we carry out (2+1)-dimensional boost-invariant hydrodynamic evolution for the same set of initial conditions, but with different equilibration timescales $\tau_\mathrm{eq}$ ranging between 0 and 10 fm/c. This evolution is carried out using a modified version of the MUSIC hydrodynamics code \cite{Schenke:2010nt, Schenke:2010rr, Paquet:2015lta}. The shear viscosities are set to match the Bayesian model averaged posterior in Ref. \cite{JETSCAPE:2020shq}, and the bulk viscosity is set to zero.

Particlization is conducted using iS3D \cite{McNelis:2019auj}, incorporating fugacity factors as detailed in the previous section. Although the distributions of hadrons at this point are modified by the quark fugacity, their dynamics afterwards are assumed to have no explicit fugacity dependence. The subsequent hadronic evolution is modeled using SMASH \cite {SMASH:2016zqf} with no modifications, until kinetic freeze-out.

\section{Numerical Results}

\subsection{Hydrodynamic Evolution}
\label{sec:evolution}

Before examining the effects of quark chemical equilibration on final particle observables, we should understand how the hydrodynamic evolution itself is impacted. We consider a single event-averaged energy density profile corresponding to a central ($b = 0$) event, and hydrodynamically evolve this same initial condition with varying $\tau_\mathrm{eq}$. 

\begin{figure}[!htbp]
    \centering
    \includegraphics[width=0.7\linewidth]{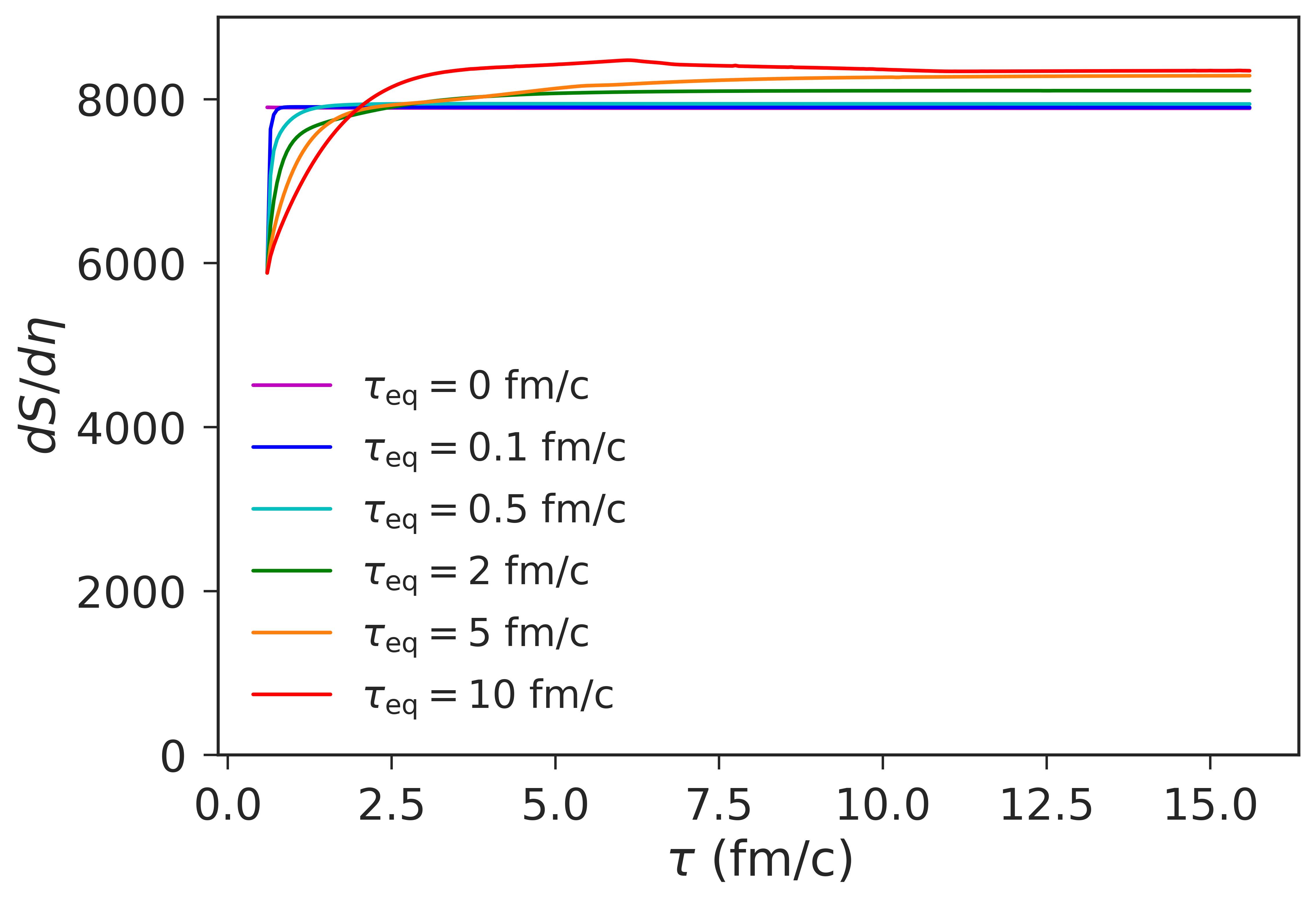}
    \caption{Total entropy per unit space-time rapidity over time for an averaged central event evolved with varying $\tau_\mathrm{eq}$ using ideal hydrodynamics.}
    \label{fig:entropy}
\end{figure}

Entropy production is one signature of quark chemical equilibration that was demonstrated in Refs. \cite{Vovchenko:2015yia,Vovchenko:2016ijt, Kurkela:2018xxd}. Entropy is conserved in ideal hydrodynamics with a fixed equation of state, but has been shown to increase when the system chemically equilibrates. We reproduce this result in Fig. \ref{fig:entropy}. Due to the initial energy density being kept constant, the total entropy starts lower for all $\tau_\mathrm{eq} > 0$ fm/c as the medium is initialized with the pure glue EoS in those cases. Depending on $\tau_\mathrm{eq}$, as much as one third of the final entropy may be produced by chemical equilibration during the hydrodynamic evolution. The entropy does not strictly increase for each timestep in Fig. \ref{fig:entropy} because it shows the entropy per unit space-time rapidity within the mid-rapidity slice of the (2+1)-dimensional computational grid, and not the total entropy of the (3+1)-dimensional physical system. The observed decrease at certain times is due to longitudinal expansion transporting matter outside the grid boundaries.

\begin{figure}[!htbp]
    \centering
    \includegraphics[width=0.7\linewidth]{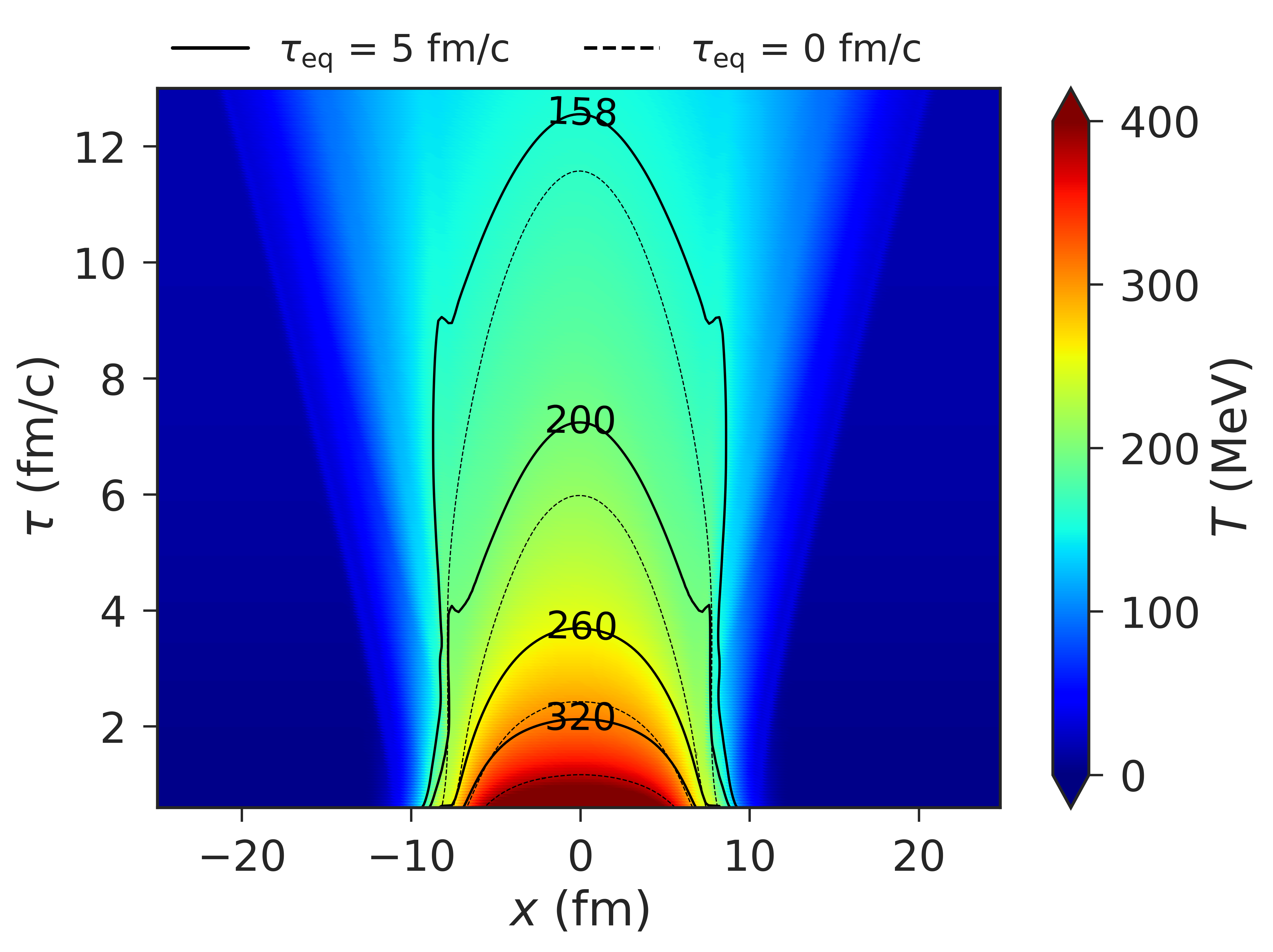}
    \caption{Contour plot of temperature in the $x-\tau$ plane for an averaged central event evolved with $\tau_\mathrm{eq} = 5$ fm/c. Solid lines show contours of $T$ in MeV, while the dashed lines show the respective isotherms when $\tau_\mathrm{eq} = 0$ fm/c.}
    \label{fig:temp_profile}
\end{figure}

Fig. \ref{fig:temp_profile} shows the temperature of the medium in the $x-\tau$ plane for $\tau_\mathrm{eq} = 5$ fm/c. In general, increasing $\tau_\mathrm{eq}$ makes the medium hotter and pushes the isotherms farther out. This is to be expected, as reducing the number of quark degrees of freedom increases the temperature at a given energy density. 

\begin{figure}[!htbp]
    \centering
    \includegraphics[width=0.7\linewidth]{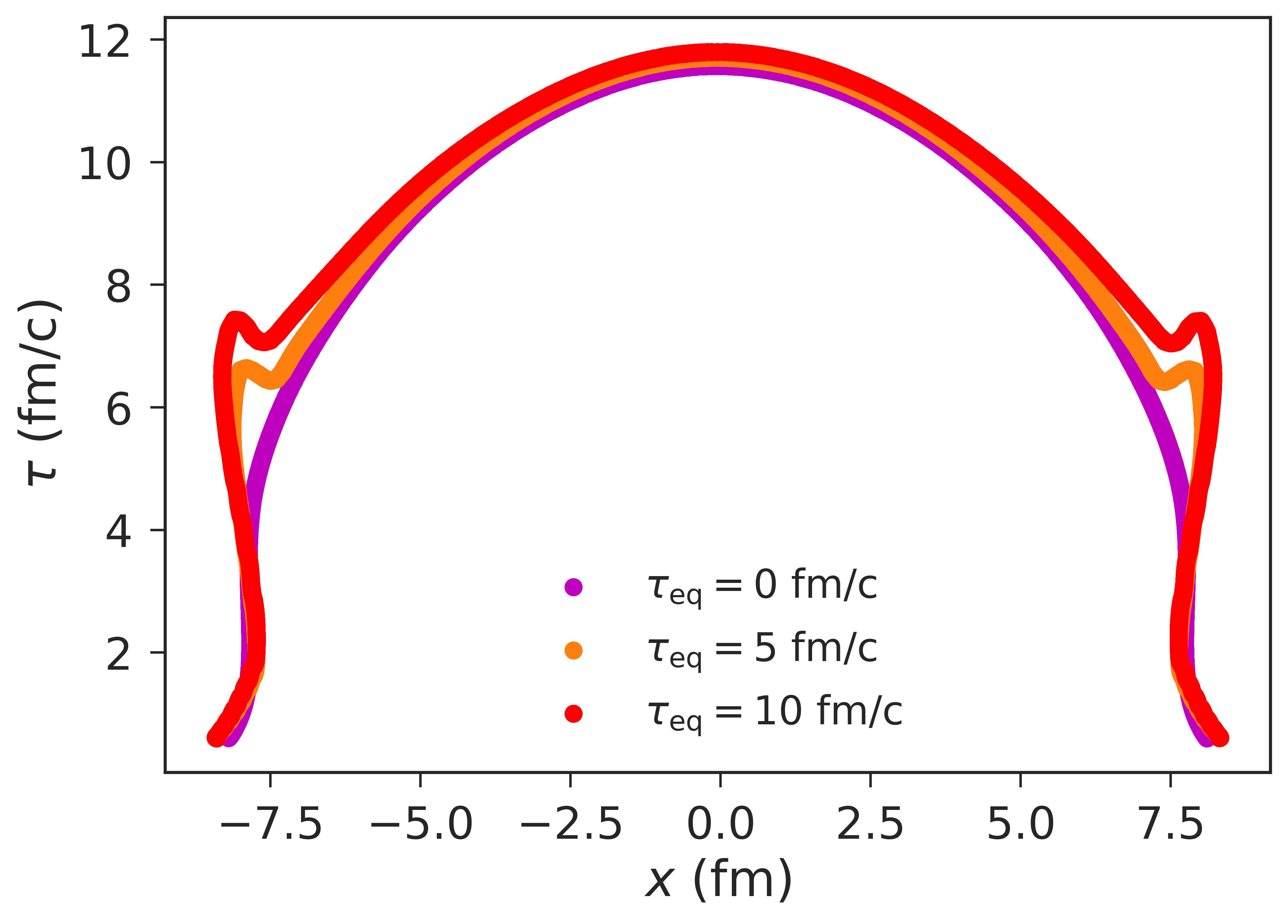}
    \caption{Particlization hypersurfaces at $T_\mathrm{c}(\gamma_q)$ in the $x-\tau$ plane for an averaged central event evolved with varying $\tau_\mathrm{eq}$.}
    \label{fig:surface_comparison}
\end{figure}

While the isotherms are notably farther from the origin with larger equilibration times, the effect is almost negligible for the particlization hypersurfaces. This is due to the choice to particlize at $T_\mathrm{c} (\gamma_q)$, which is higher for lower fugacities. Thus, when $\tau_\mathrm{eq}$ is larger, we particlize across a surface with higher temperature and reduced quark content, but the volume and energy density changes little. Fig. \ref{fig:surface_comparison} demonstrates this by comparing the hypersurfaces for several values of $\tau_\mathrm{eq}$. 

\begin{figure}[!htbp]
    \centering
    \includegraphics[width=0.7\linewidth]{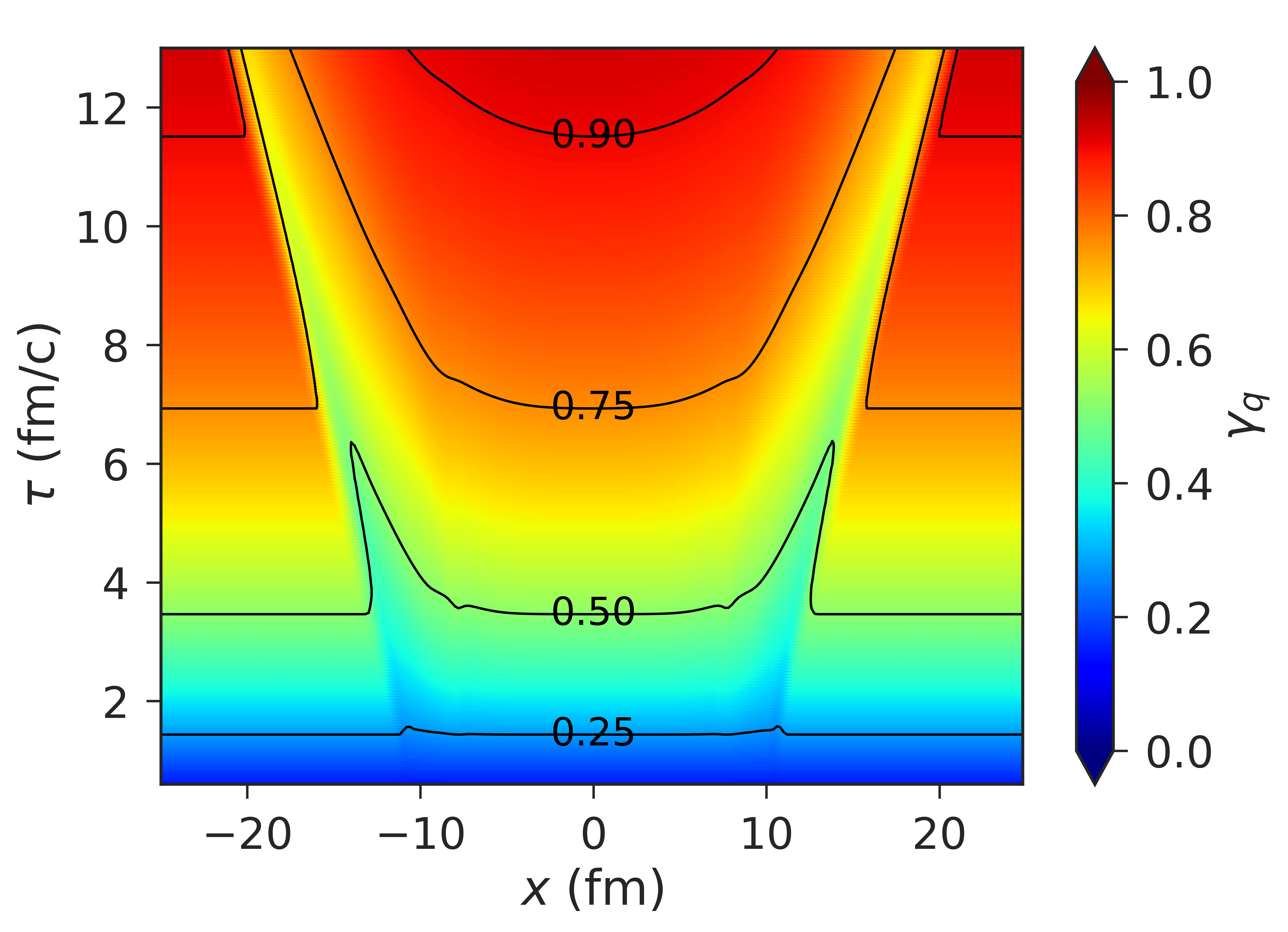}
    \caption{Contour plot of fugacity in the $x-\tau$ plane for an averaged central event evolved with $\tau_\mathrm{eq} = 5$ fm/c.}
    \label{fig:fugacity_profile}
\end{figure}

Fig. \ref{fig:fugacity_profile} shows the fugacity in the $x-\tau$ plane for the same event at $\tau_\mathrm{eq} = 5$ fm/c. As expected from the proper time dependence of $\gamma_q$ in Eq. \ref{eq:fugacity}, cells nearer the periphery of the fluid equilibrate more slowly due to their greater velocities. Note that for large enough $x$, where there is effectively zero energy density, the local proper time $\tau_\mathrm{p}$ is equivalent to the global simulation time $\tau$. This is irrelevant in practice, as this dilute region is well outside the particlization hypersurface.

\begin{figure}[!htbp]
    \centering
    \includegraphics[width=0.7\linewidth]{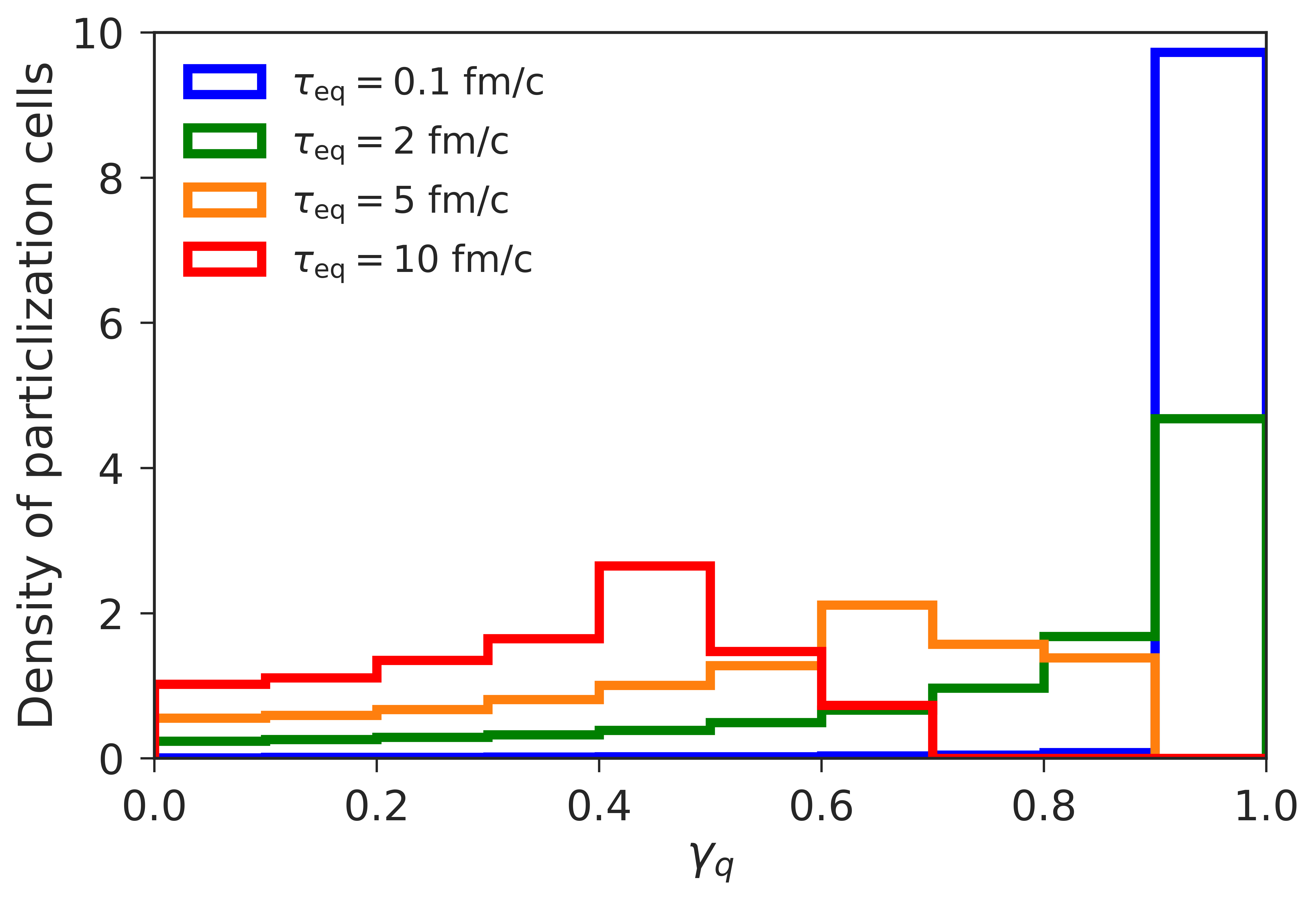}
    \caption{Fugacity distribution of cells in the particlization hypersurface for an averaged central event evolved with varying $\tau_\mathrm{eq}$.}
    \label{fig:fugacity_density}
\end{figure}

It is also worth noting that the entire surface at $\tau_\mathrm{eq} = 5$ fm/c is within the region where $\gamma_q < 0.9$, and much of it falls within $\gamma_q < 0.5$. To show this more clearly, Fig. \ref{fig:fugacity_density} shows the fugacity distribution of cells in the surface for various $\tau_\mathrm{eq}$. Unsurprisingly, selecting a small enough equilibration timescale produces a surface that is almost entirely equilibrated. For large enough $\tau_\mathrm{eq}$, on the other hand, the entire medium can particlize far from equilibrium - as low as $\gamma_q < 0.7$ for the largest timescale we consider, $\tau_\mathrm{eq} = 10$ fm/c.

\begin{figure}[!htbp]
    \centering
    \includegraphics[width=0.7\linewidth]{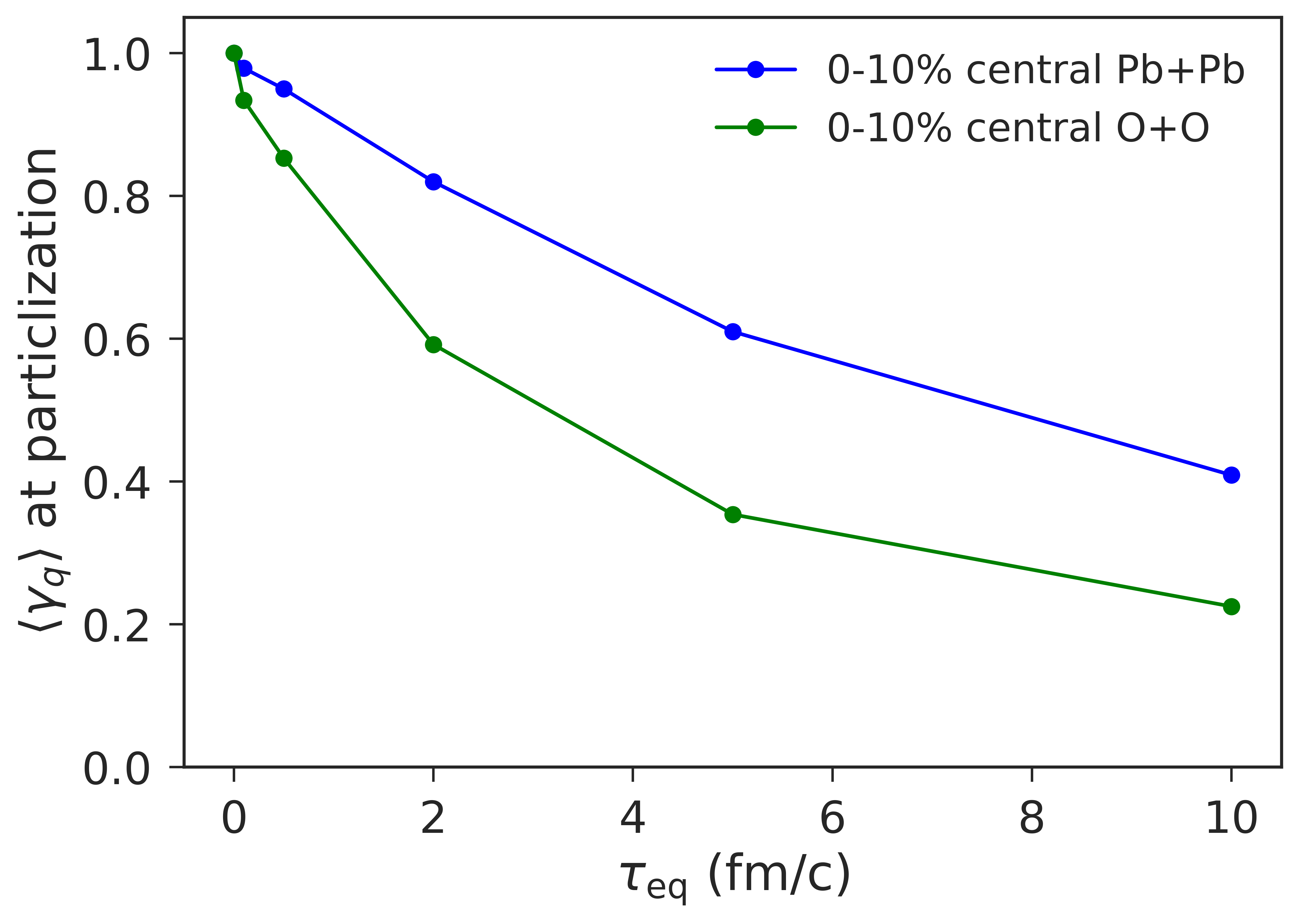}
    \caption{Mean fugacity of the particlization hypersurface for averaged central Pb+Pb and O+O events evolved with varying $\tau_\mathrm{eq}$.}
    \label{fig:fugacity_means}
\end{figure}

\begin{figure}[!htbp]
    \centering
    \includegraphics[width=0.7\linewidth]{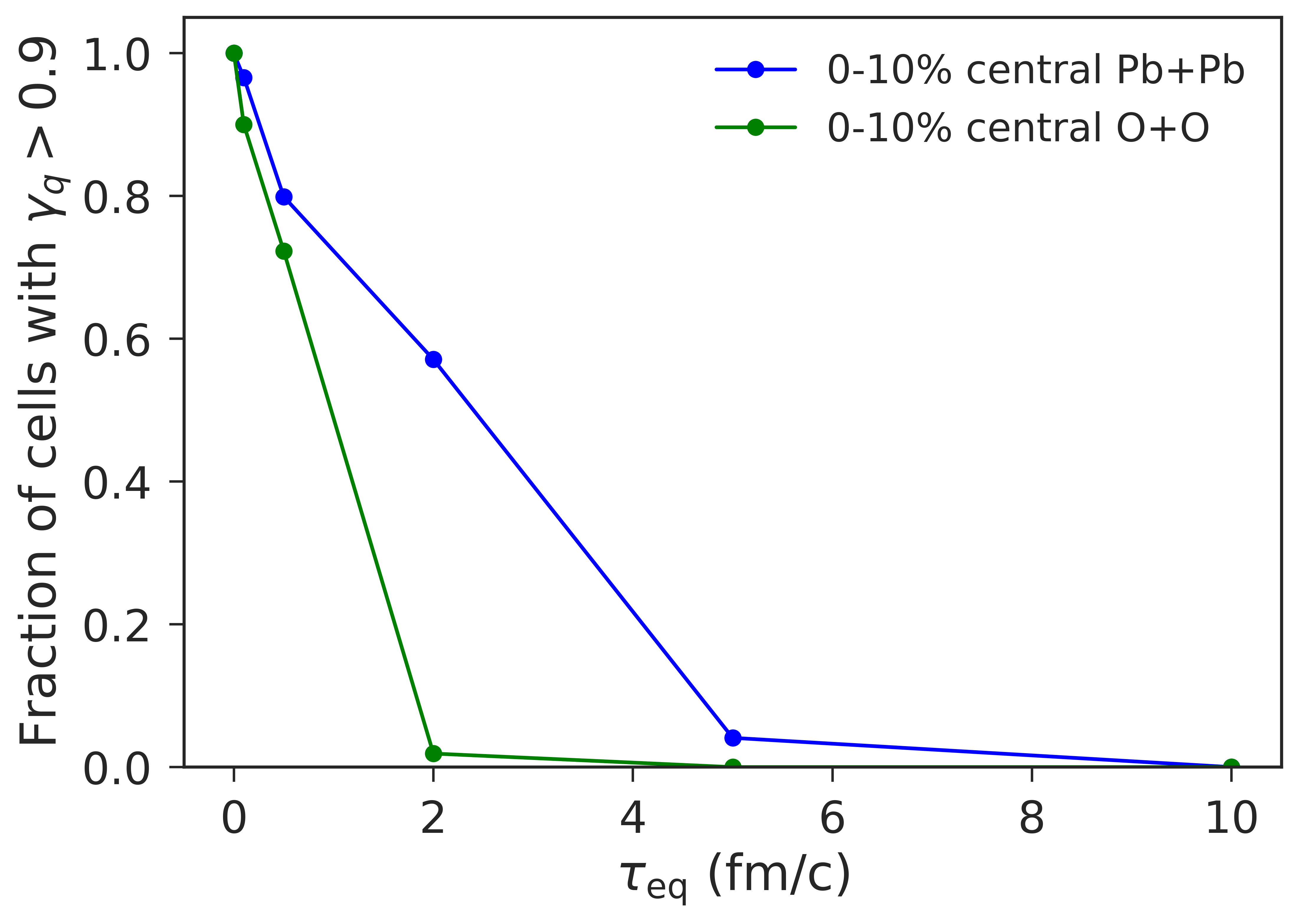}
    \caption{Proportion of particlization hypersurface cells near chemical equilibrium for averaged central Pb+Pb and O+O events evolved with varying $\tau_\mathrm{eq}$.}
    \label{fig:fugacity_0p9}
\end{figure}

In this work, we primary consider Pb+Pb events, which are among the largest nuclei collided experimentally. However, the model being presented is general enough to apply to any collision sufficiently large to produce a QGP droplet. Smaller systems will tend to have less time to chemically equilibrate before hadronization, and we expect this to be reflected in the quark fugacity at particlization. To demonstrate this effect, we compare the particlization hypersurfaces produced for the aforementioned Pb+Pb profile with those for an initial condition corresponding to a single event-averaged central O+O collision. Fig. \ref{fig:fugacity_means} shows the resulting mean fugacities as a function of $\tau_\mathrm{eq}$.  One can see that for any given $\tau_\mathrm{eq} > 0,$ the mean fugacity is consistently lower for O+O, as expected. Additionally, Fig. \ref{fig:fugacity_0p9} shows the proportion of fluid cells in the hypersurface with a fugacity of $\gamma_q > 0.9,$ which one can interpret as roughly the fraction of cells at or near chemical equilibrium. In this light, it is apparent that even for a moderate timescale of $\tau_\mathrm{eq} = 2$ fm/c, almost none of the system is near chemical equilibrium in the case of an O+O collision.

\subsection{Hadron Production}

It is now apparent that for larger equilibration timescales, particlization occurs at a higher temperature and at lower quark fugacities. This leads to two competing effects on hadron production: Higher temperatures at particlization generally increase the yields of hadrons, as can be seen from their distribution functions in Eq. \ref{eq:f_i}. Conversely, lower fugacities correspond to larger suppression factors in the hadron distribution functions. The net effect on hadron production is thus determined by the interplay between these two factors.

\begin{figure}[!htbp]
    \centering
    \includegraphics[width=0.7\linewidth]{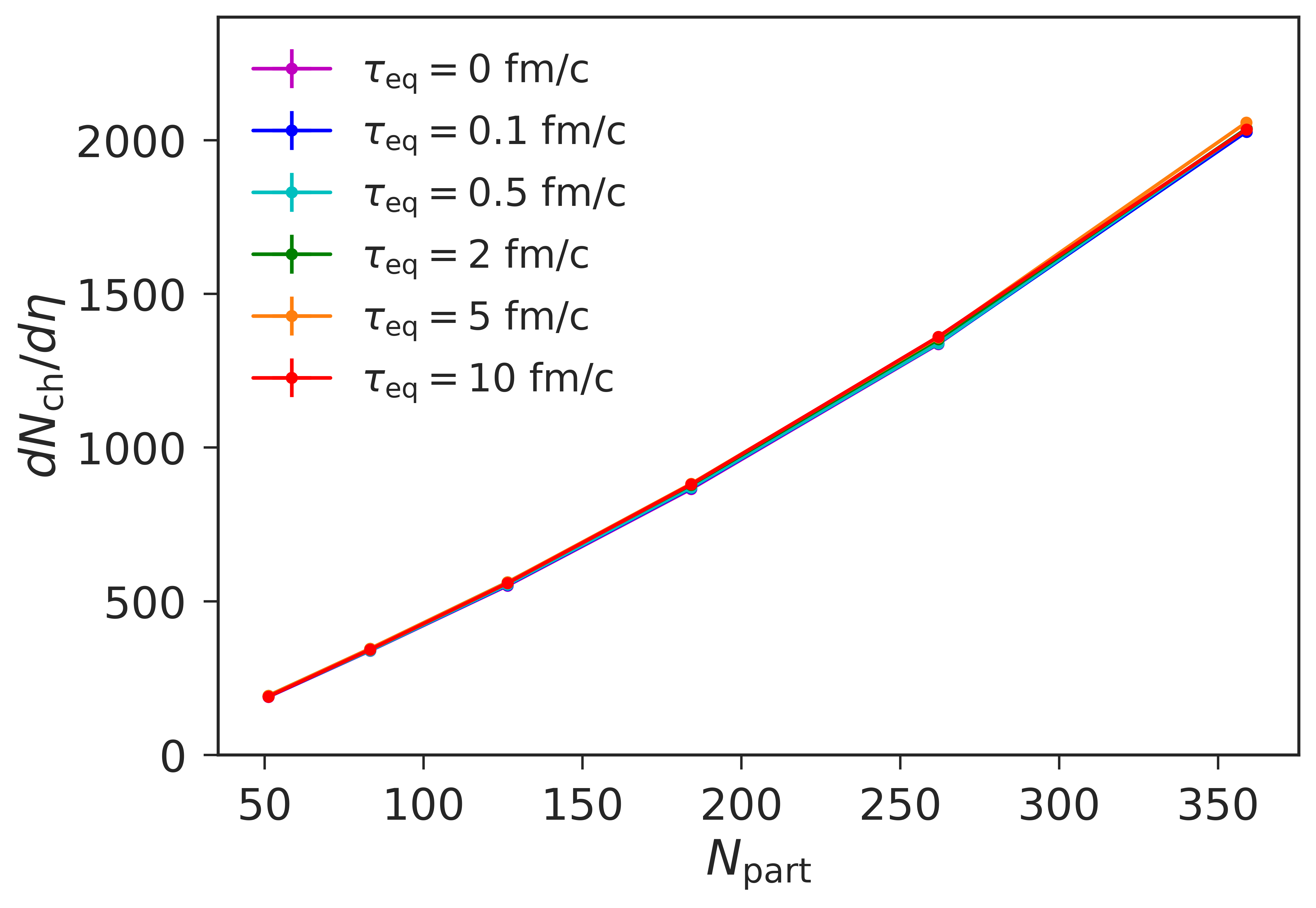}
    \caption{Charged particle multiplicity for 0-60\% centrality events evolved with varying $\tau_\mathrm{eq}$. Each point corresponds to a 10\% centrality bin.}
    \label{fig:dNch}
\end{figure}

For the following results, we consider an ensemble of 10,000 minimum bias events, each evolved at several values of $\tau_\mathrm{eq}$. All observables are calculated at midrapidity with the cut $|\eta| < 0.5$, and the events are binned by centrality in increments of 10\% according to their total initial entropy. Fig. \ref{fig:dNch} shows the resulting charged particle multiplicity for the six most central bins (0-60\%), plotted against the number of participant nucleons in the initial collision, $N_{part}.$ Surprisingly, when varying $\tau_\mathrm{eq}$ from 0 to 10 fm/c, the multiplicities consistently agree within $\approx 2\%.$ This suggests that there is a close cancellation between the effects due to higher temperatures and lower fugacities at particlization.

\begin{figure*}[htbp]
    \centering
    \includegraphics[width=0.49\textwidth]{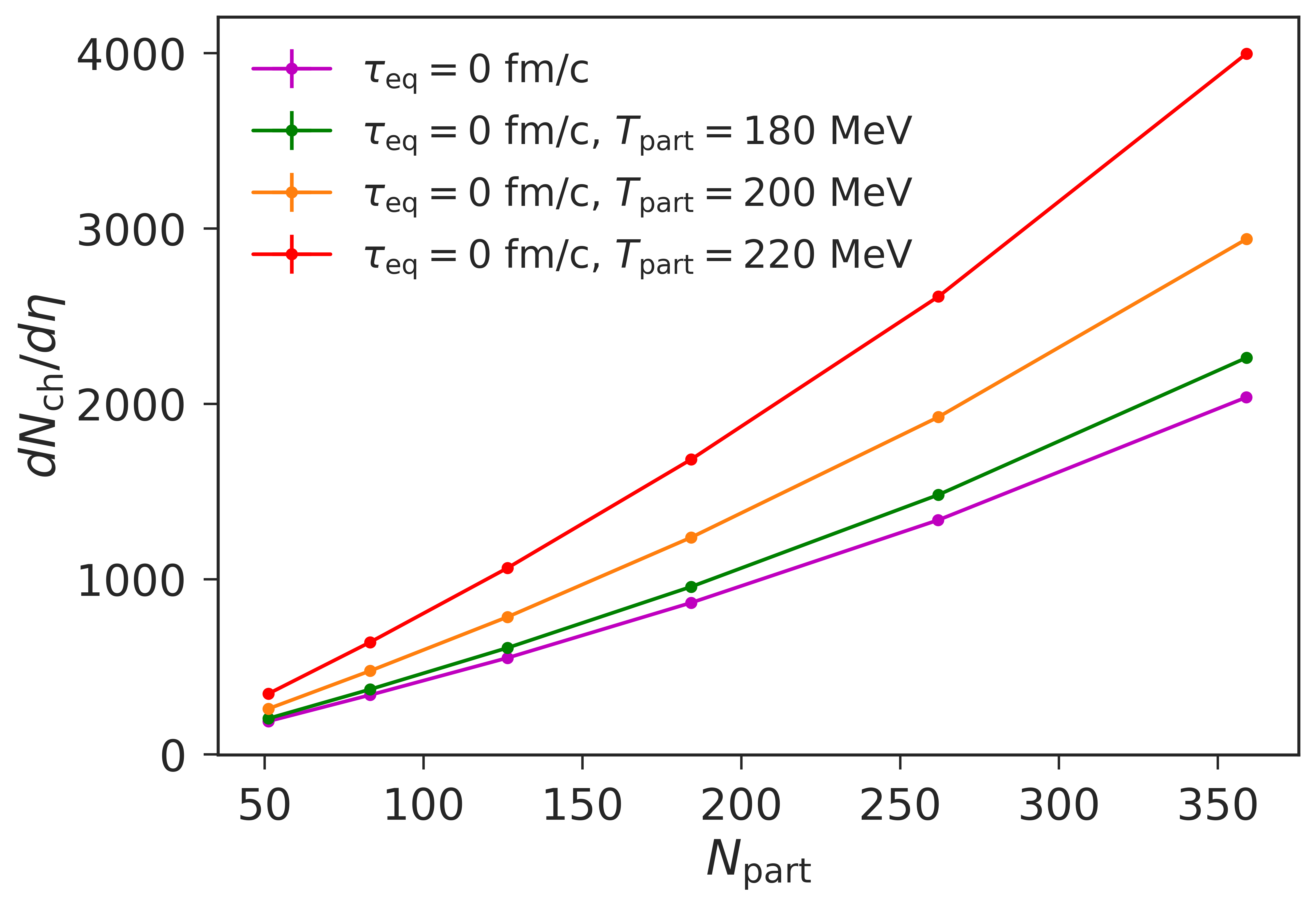}%
    \hfill
    \includegraphics[width=0.49\textwidth]{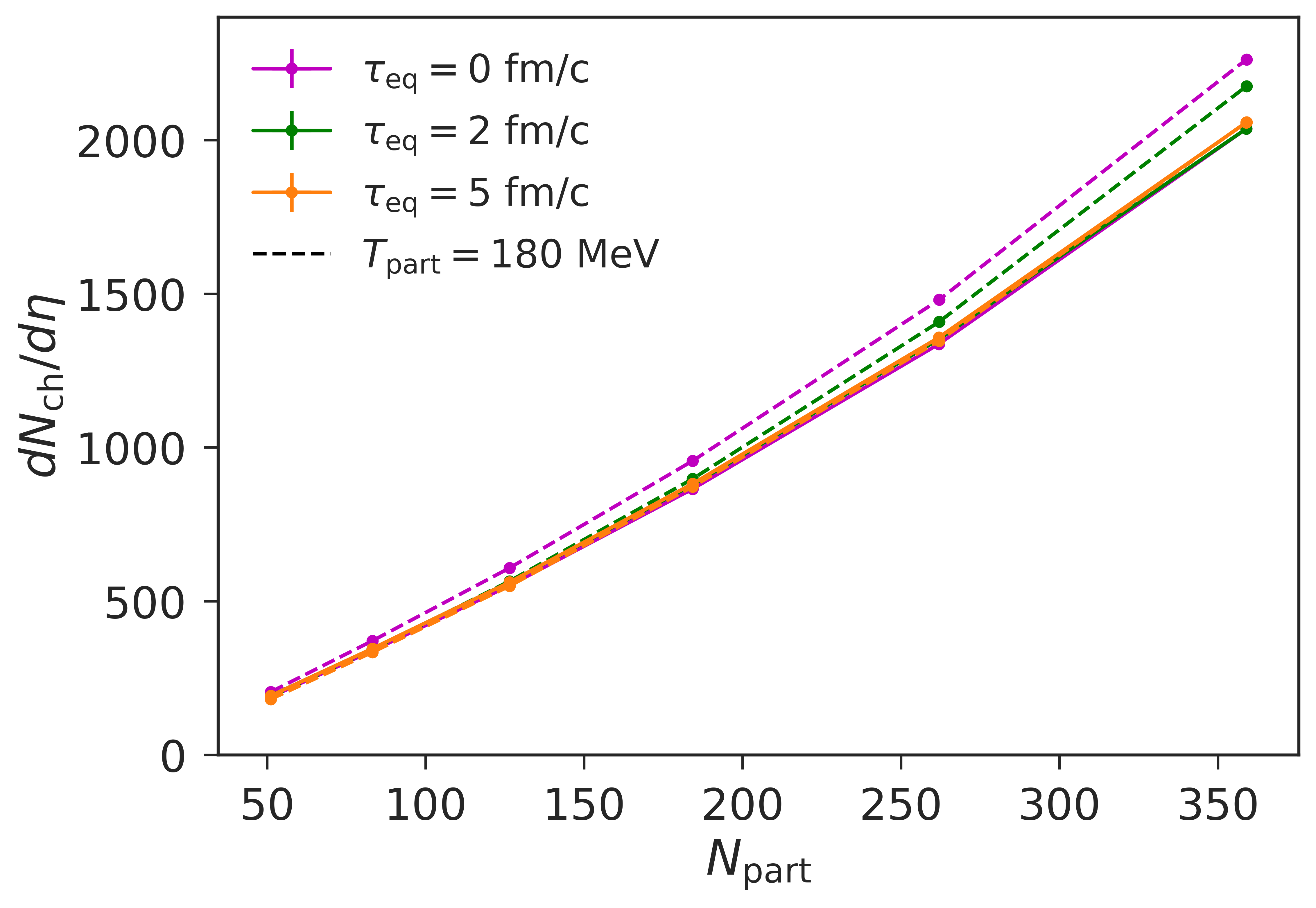}
    \caption{Charged particle multiplicity for 0--60\% centrality events evolved with varying particlization temperature $T_{\text{part}}$ (left) or varying $\tau_{\text{eq}}$ at a fixed $T_{\text{part}}$ (right). The curves are color-coded such that the value of $T_{\text{part}}$ for each curve on the left is approximately the mean particlization temperature for the corresponding colored solid curve on the right (e.g., $\langle T_{\text{part}} \rangle \approx 200~\text{MeV}$ for $\tau_{\text{eq}} = 5~\text{fm}/c$). Each point corresponds to a 10\% centrality bin.}
    \label{fig:dNch_temperature}
\end{figure*}

Fig. \ref{fig:dNch_temperature} shows the effects due to temperature and fugacity individually on the charged particle multiplicities. It is clear from the left plot that increasing the particlization temperature substantially increases the multiplicity, by as much as a factor of two. At the same time, the right plot demonstrates that at a fixed particlization temperature, increasing the equilibration timescale reduces the multiplicity, albeit not as drastically.

\begin{figure*}[htbp]
    \centering
    \includegraphics[width=0.49\textwidth]{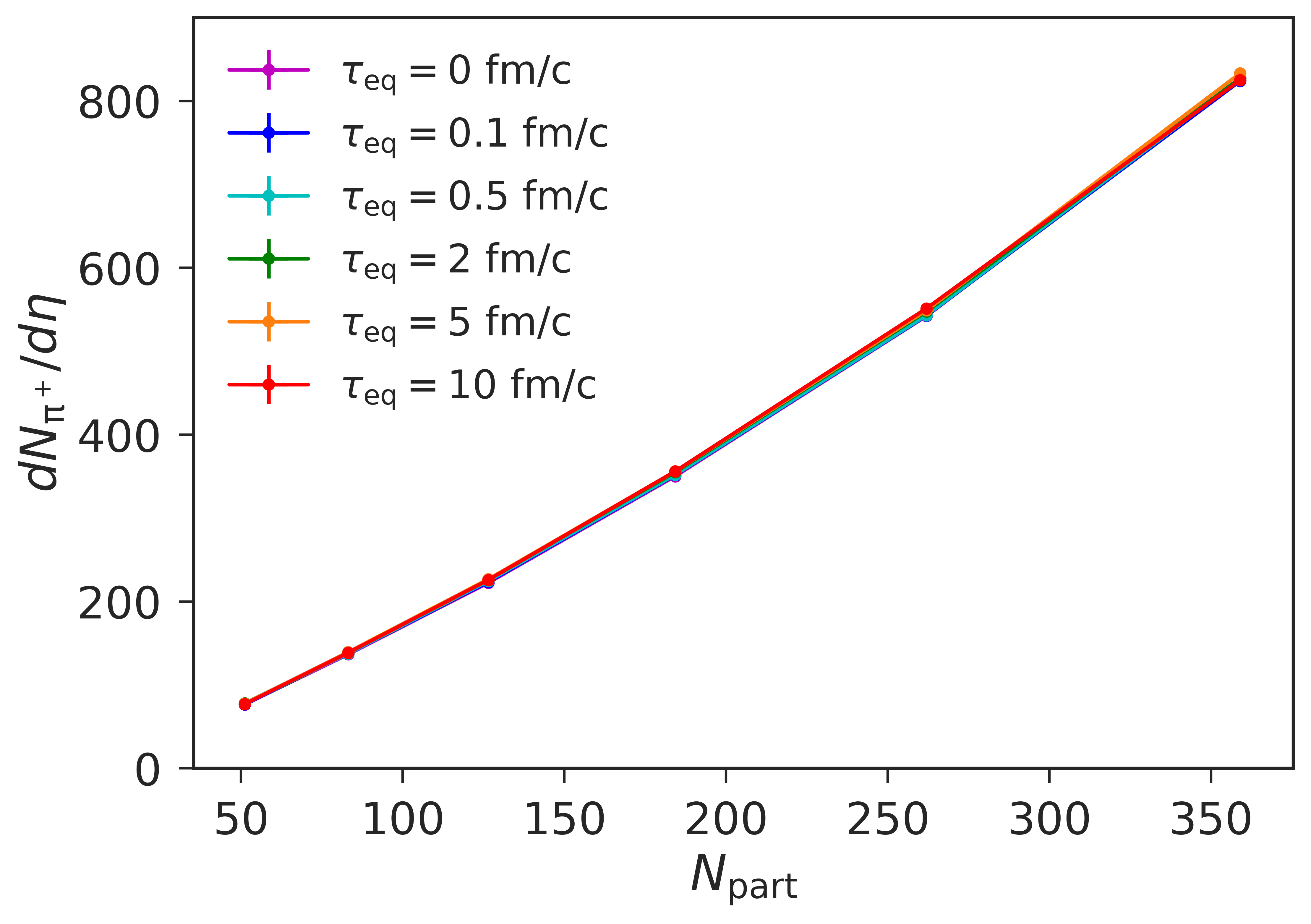}%
    \hfill
    \includegraphics[width=0.49\textwidth]{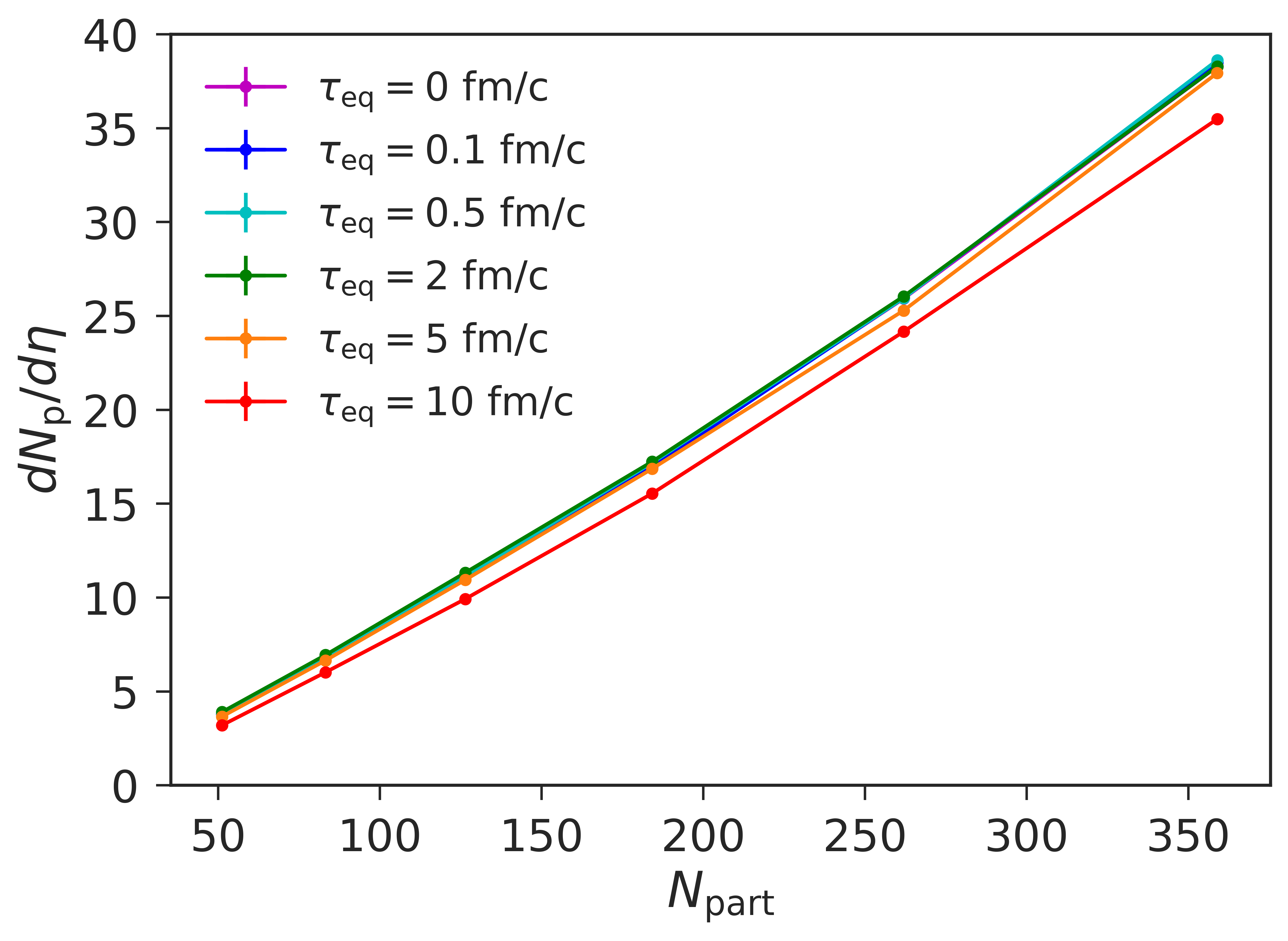}
    \caption{Pion (left) and proton (right) multiplicities for 0-60\% centrality events evolved with varying $\tau_\mathrm{eq}$. Each point corresponds to a 10\% centrality bin.}
    \label{fig:dNppi}
\end{figure*}

\begin{figure}[!hbtp]
    \centering
    \includegraphics[width=0.7\linewidth]{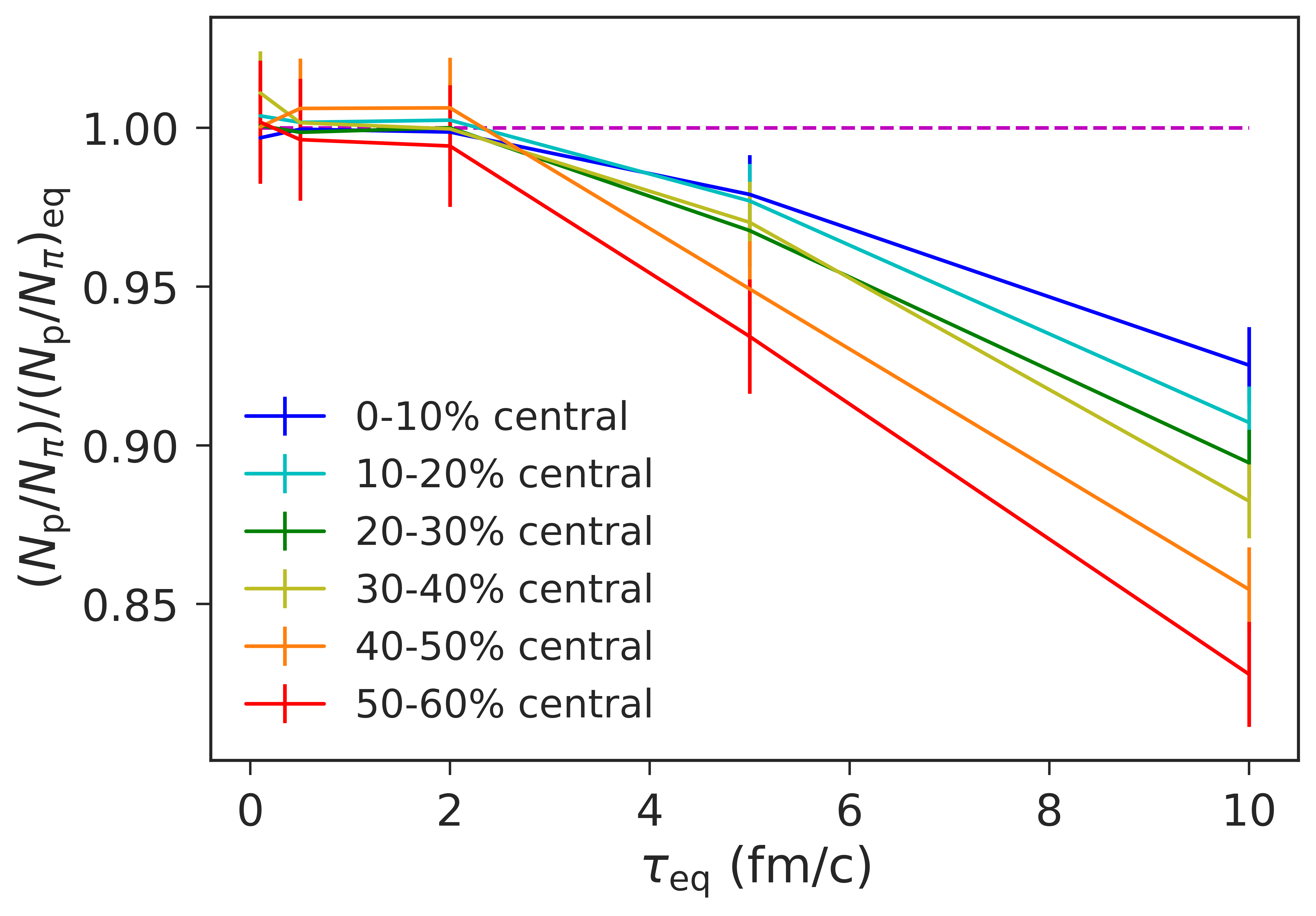}
    \caption{Modification to pion-to-proton ratio for 0-60\% centrality events evolved with varying $\tau_\mathrm{eq}$, as compared to this ratio with $\tau_\mathrm{eq} = 0$ fm/c.}
    \label{fig:p_pi}
\end{figure}

Baryon suppression by a factor of $\approx 1.5$ relative to expectations from thermal models has been observed experimentally in $\sqrt{s_{NN}} = 2.76 $ TeV Pb+Pb collisions at the LHC \cite{ALICE:2013mez}. This has been understood as a consequence of baryon-antibaryon annihilation into pions \cite{Steinheimer:2012rd, Karpenko:2012yf, Becattini:2012xb}, but it has also been hypothesized that chemical undersaturation of the QGP may contribute, particularly for peripheral collisions \cite{Vovchenko:2015yia}. This is evident in our model, as the fugacity factors assigned to the hadron distribution functions as defined in Eq. \ref{eq:lambda_i} naturally suppress baryons more than mesons. Fig. \ref{fig:dNppi} shows the pion and proton multiplicities produced by our model, while Fig. \ref{fig:p_pi} shows the modification to the pion-to-proton ratio as a function of $\tau_\mathrm{eq}$. For equilibration times up to $\tau_\mathrm{eq} = 2$ fm/c, the variation in both the pion and proton multiplicities is negligible. Only for $\tau_\mathrm{eq} \geq 5$ fm/c do we observe the proton multiplicity and proton-to-pion ratio decrease, providing a weak signal of baryon suppression. However, this is a $\lesssim 20$\% effect that does not explain the much larger suppression of the $p/\pi$ ratio seen in experiments, and instead should be considered as a correction in conjunction with other models of baryon suppression.

\subsection{Transverse Flow}

\begin{figure}[!htbp]
    \centering
    \includegraphics[width=0.7\linewidth]{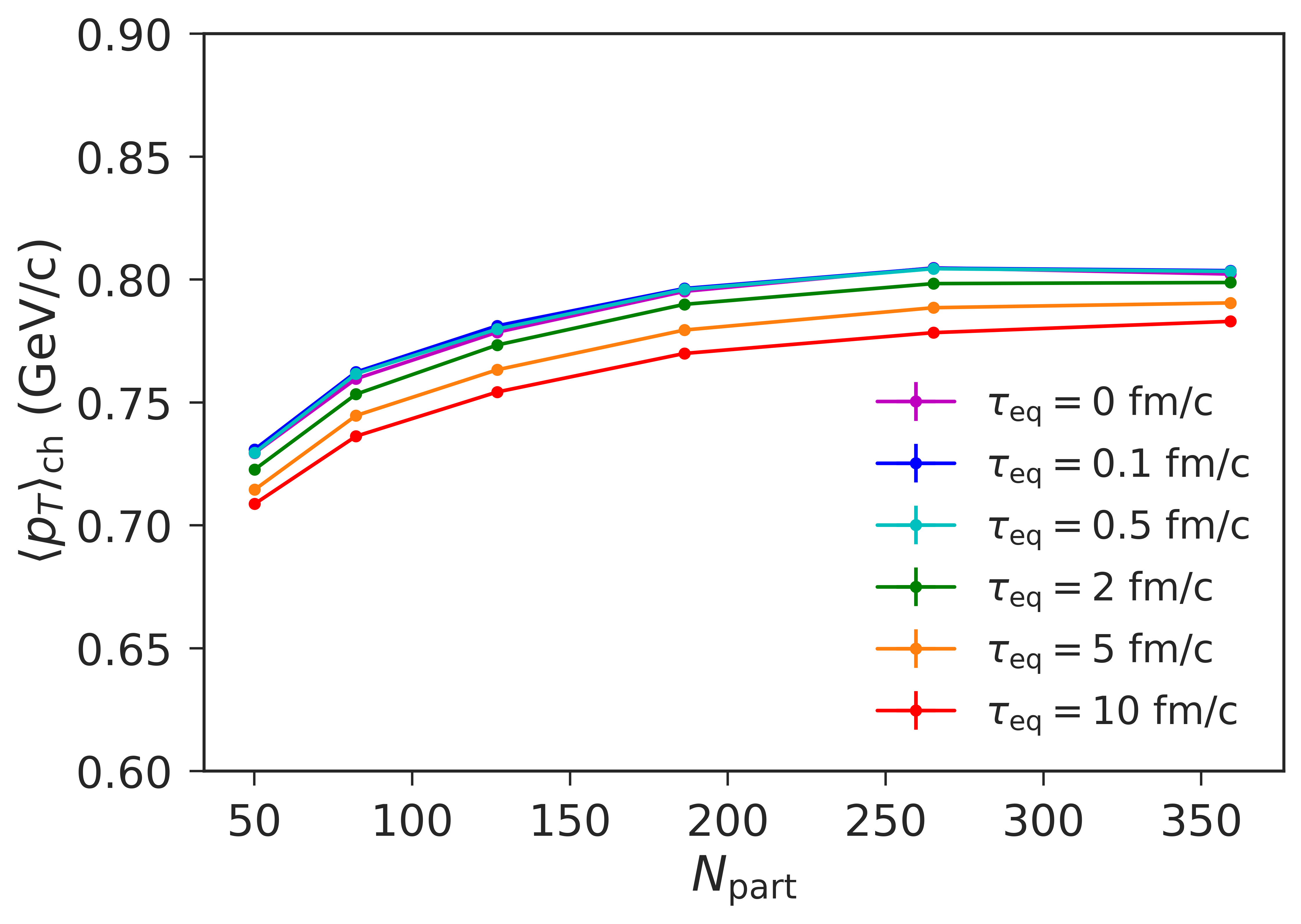}
    \caption{Mean transverse momentum of charged particles for 0-60\% centrality events evolved with varying $\tau_\mathrm{eq}$. Each point corresponds to a 10\% centrality bin.}
    \label{fig:pTch}
\end{figure}

\begin{figure*}[htbp]
    \centering
    \includegraphics[width=0.49\textwidth]{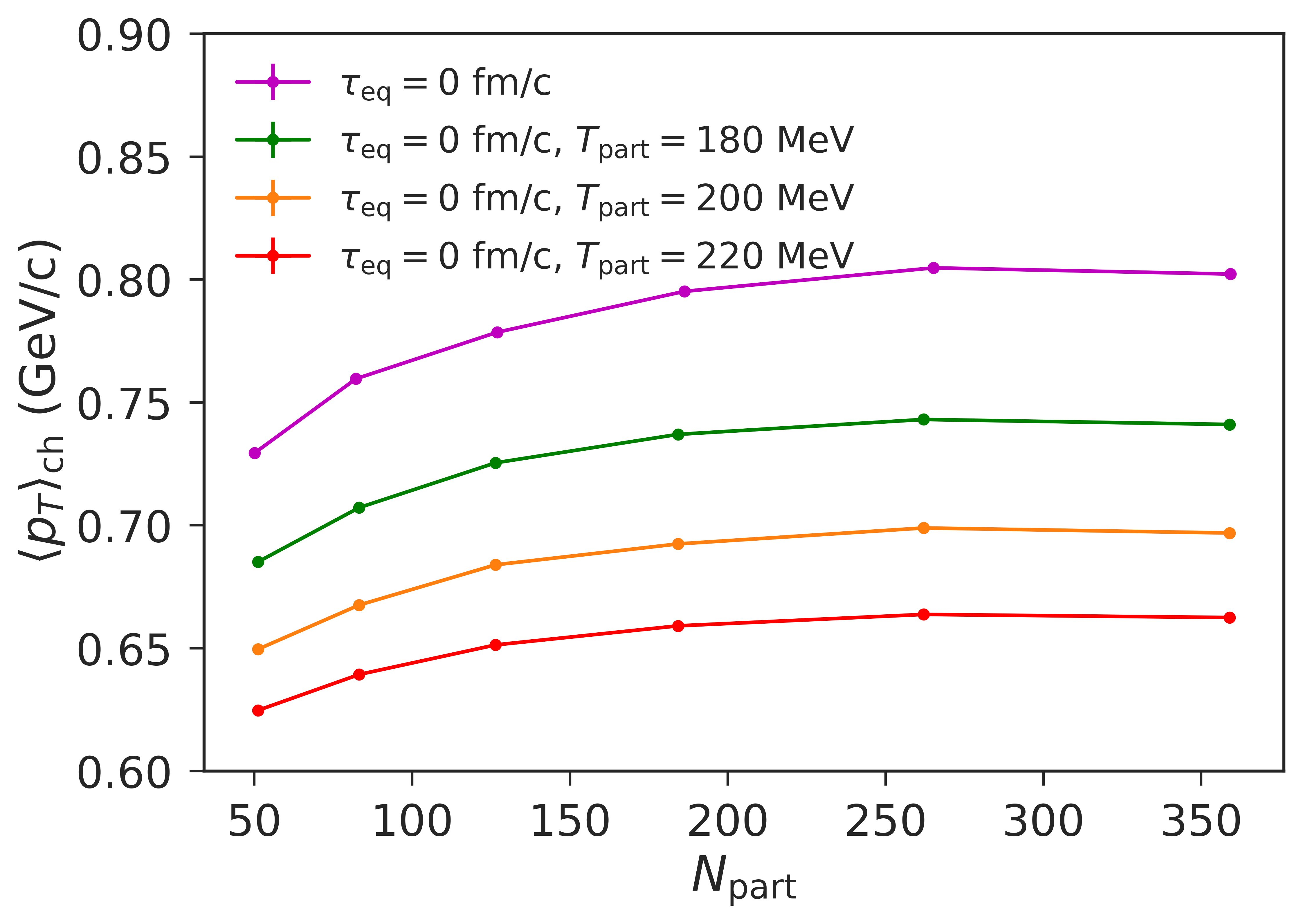}%
    \hfill
    \includegraphics[width=0.49\textwidth]{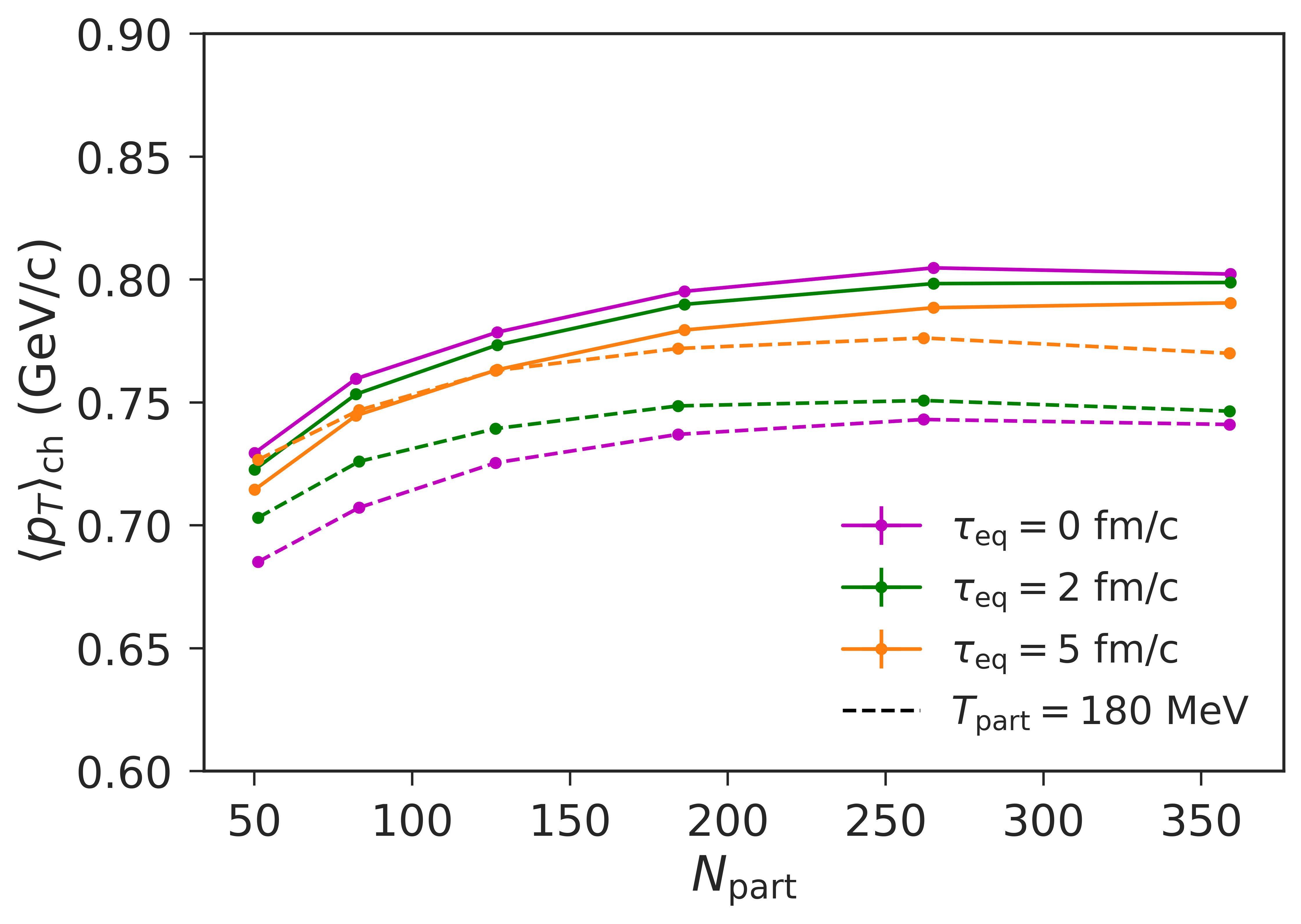}
    \caption{Mean transverse momentum of charged particles for 0-60\% centrality events evolved with varying particlization temperature $T_\mathrm{part}$ (left) or varying $\tau_\mathrm{eq}$ at a fixed $T_\mathrm{part}$(right). The curves are color-coded such the value of $T_\mathrm{part}$ for each curve on the left is approximately the mean particlization temperature for the corresponding colored solid curve on the right (e.g., $\langle T_\mathrm{part} \rangle \approx 200$ MeV for $\tau_\mathrm{eq} = 5$ fm/c). Each point corresponds to a 10\% centrality bin.}
    \label{fig:pTch_temperature}
\end{figure*}

Quark chemical equilibration can affect not only the production of hadrons at particlization, but also the development of flow during the evolution of the QGP. This is ultimately reflected in the momenta of the final state hadrons. Fig. \ref{fig:pTch} shows the mean transverse momentum $\langle p_T \rangle$ of charged particles, while Fig. \ref{fig:pTch_temperature} shows the independent effects of the particlization temperature and fugacity. 

As $\tau_\mathrm{eq}$ decreases, the mean transverse momentum systematically decreases. In chemical equilibrium, increasing the particlization temperature reduces the evolution time and thus allows less flow to develop, as in Fig. \ref{fig:dNch_temperature}. However, as shown in Fig. \ref{fig:surface_comparison}, the evolution time does not decrease with $\tau_\mathrm{eq}$. In fact, when $\tau_\mathrm{eq}$ is increased and the particlization temperature is fixed, as in Fig. \ref{fig:dNch_temperature}, the evolution time and thus mean $\langle p_T \rangle$ increase. The decrease in $\langle p_T \rangle$ should instead be attributed to the reduced pressure out of equilibrium, as lower quark fugacities correspond to lower $P$ throughout the evolution and it is pressure gradients that drive the development of transverse flow. 

\begin{figure}[!htbp]
    \centering
    \includegraphics[width=0.7\linewidth]{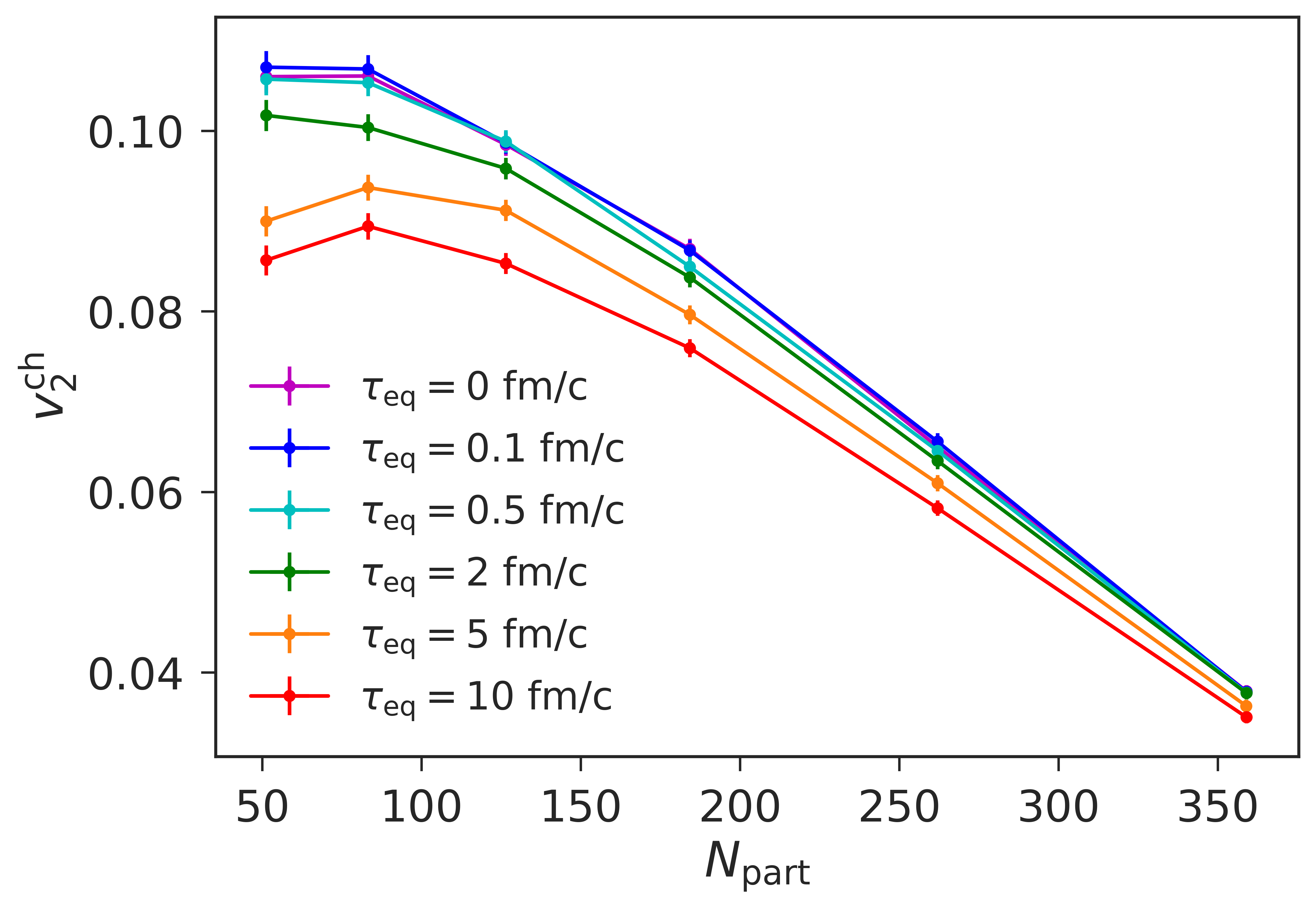}
    \caption{Elliptic flow for 0-60\% centrality events evolved with varying $\tau_\mathrm{eq}$. Each point corresponds to a 10\% centrality bin.}
    \label{fig:v2ch}
\end{figure}

A similar effect can be observed with the flow anisotropy. Transverse momentum anisotropy is commonly expressed in terms of the coefficients $v_n$ of the Fourier expansion

\begin{align}
    \frac{dN}{d\phi} \propto 1 + 2 \sum_{n=1}^\infty v_n \cos{n(\phi - \Phi_n)},
\end{align}

where $\phi = \text{atan2}(p_y,p_x)$ is the azimuthal angle and $\Phi_n$ is the event-plane angle. The elliptic flow $v_2$ is particularly useful as a measure of the conversion from elliptical anisotropy in the initial condition to momentum anisotropy. For the ensemble of simulated events, the $v_2$ of charged particles, is shown in Fig. \ref{fig:v2ch}. The suppression of $v_2$ due to quark chemical equilibration is greater than that of $\langle p_T \rangle$, and this is particularly noticeable for less central events. This is to be expected, as the smaller size and shorter lifetime of less central events amplify the effects of weaker pressure gradients. It is known that $v_2$ tends to decrease with greater shear and bulk viscosities \cite{Song:2007ux, Shen:2011kn, Dusling:2011fd, Ryu:2017qzn}, so one can interpret quark chemical equilibration as effectively increasing these viscosities. In this work, we did not vary $\Pi$ and $\pi^{\mu \nu}$ with the quark fugacity, but this result highlights the necessity of doing so in future studies. When modeling the QCD medium in partial chemical equilibrium, smaller viscosities may yield better agreement with experimental data due to the effective viscosity increase implied by quark chemical equilibration.

\subsection{Thermal Photon Production}

Photons are emitted from various sources throughout the stages of a heavy-ion collision: prompt photons are produced from nucleon interactions in the initial stages, thermal photons are emitted by the thermalized medium as it expands, and additional photons are produced by hadronic decays. We focus on thermal photons here as the most direct probe of the deconfined phase.

To leading order, there are four processes by which thermal photons are produced in the QGP: gluon-photon Compton scattering, elastic quark-antiquark annihilation, bremsstrahlung, and inelastic pair annihilation \cite{Arnold:2001ms}. With a medium in partial chemical equilibrium, the photon production rate for each process should be suppressed by a factor of $\gamma_q$ for each (anti)quark. Gluon-photon Compton scattering ($q/\bar{q} + g \rightarrow q/\bar{q} + \gamma$) is then suppressed linearly by $\gamma_q$, and elastic pair annihilation ($q + \bar{q} \rightarrow g + \gamma$) is suppressed quadratically by $\gamma_q^2$. It is more difficult to determine the correct suppression for the two inelastic processes. Bremsstrahlung may occur with either two (anti)quarks or one (anti)quark and one gluon, and inelastic pair production involves scattering of a (anti)quark on another parton that may be either a (anti)quark or gluon. 

Adopting the scheme of Ref. \cite{Vovchenko:2016ijt}, we compare two approximations in which the inelastic processes are suppressed either linearly or quadratically with respect to $\gamma_q$. Defining $\Gamma (k,T,\gamma_q)$ as the production rate for photons with energy $k$ in a fluid cell with temperature $T$ and fugacity $\gamma_q$, we have: 

\begin{equation}
\begin{aligned}
\Gamma(k,T,\gamma_q) = \gamma_q\, \Gamma_{\mathrm{Compton}}(k,T)
 + \gamma_q^2\, \Gamma_{\mathrm{annihilation}} + \gamma_q^n\, \Gamma_{\mathrm{inelastic}},
\end{aligned}
\end{equation}

where $n \in (1,2)$, and $\Gamma_{\text{Compton}}, \Gamma_{\text{annihilation}}, $ and $\Gamma_{\text{inelastic}}$ are the respective photon production rates of gluon-photon Compton scattering, elastic pair annihilation, and inelastic processes. $n = 1$ corresponds to a scheme that underestimates the total thermal photon production rate, and $n = 2$ corresponds to a scheme that overestimates it. Using the expressions for these three rates given in Ref. \cite{Arnold:2001ms}, we calculate thermal photon emission for the single averaged central event evolved at varying $\tau_\mathrm{eq}$ referred to in Sec. \ref{sec:evolution}. We also neglect viscous corrections here, and only consider the ideal photon production rate.

\begin{figure}[!htbp]
    \centering
    \includegraphics[width=0.7\linewidth]{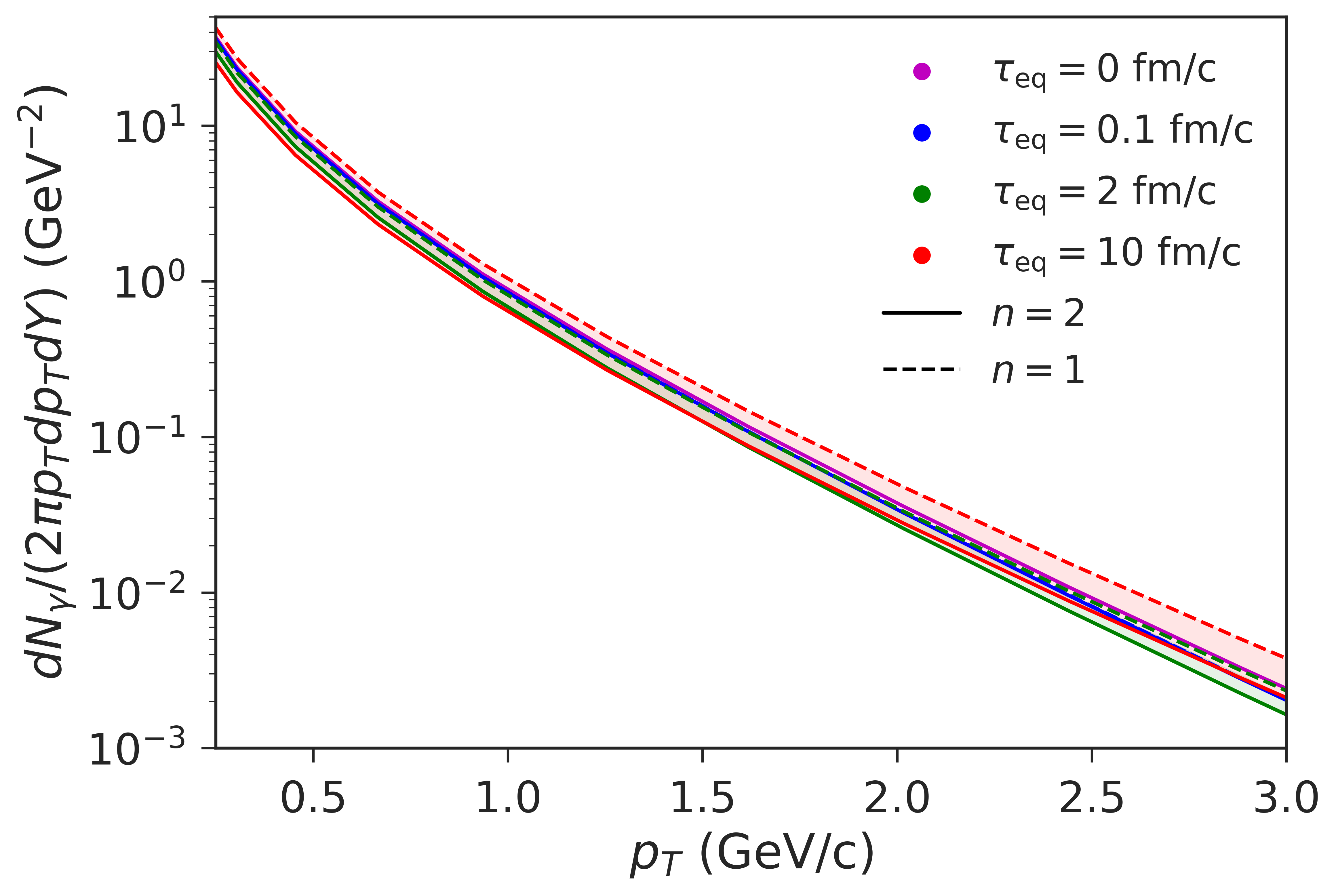}
    \includegraphics[width=0.7\linewidth]{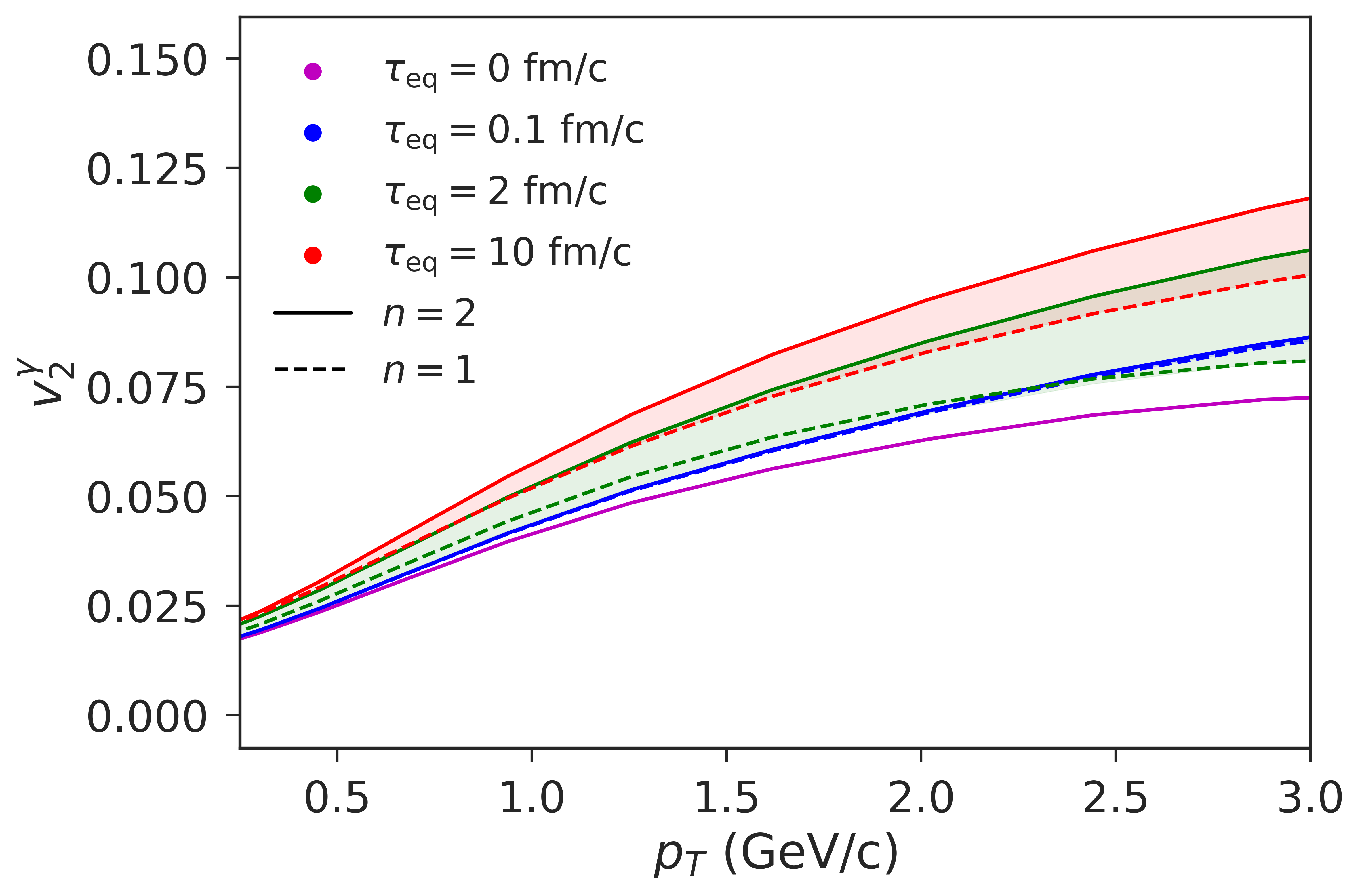}
    \caption{Thermal photon spectrum (top) and elliptic flow (bottom) for an averaged central event evolved with varying $\tau_\mathrm{eq}$, using a cutoff temperature of $150$ MeV. Solid lines correspond to quadratic scaling of $\Gamma_{\text{inelastic}}$ with respect to $\gamma_q$ and dashed lines correspond to linear scaling.}
    \label{fig:photons}
\end{figure}

For (2+1)-dimensional boost-invariant hydrodynamics, the $p_T$ spectrum of thermal photons is given by

\begin{equation}
\frac{dN_\gamma}{d^2p_T\,dy}
= \int d^2x_T d\tau d\eta_s \tau
  \Gamma(k,T,\gamma_q)\,
  \theta\!\left(T - T_{\mathrm{min}}\right),
\end{equation}

where $\theta(x)$ is the Heaviside step function and $T_{\text{min}}$ is the minimum temperature for emission, here set to $150$ MeV. In other words, we neglect late stage photon emission from fluid cells with $T < 150$ MeV. Fig. \ref{fig:photons} shows the resulting spectrum together with the elliptic flow of thermal photons, which is given by

\begin{align}
    v_2^\gamma (p_T) = \frac{\int_0^{2\pi} d\phi \frac{dN_\gamma}{d^2p_Tdy} \cos(2\phi)}{\int_0^{2\pi} d\phi \frac{dN_\gamma}{d^2p_Tdy}}.
\end{align}

The results in Fig. \ref{fig:photons} show that both the spectra and elliptic flow of thermal photons are sensitive to the quark chemical equilibration timescale. Most notably, $v_2^\gamma$ tends to be enhanced at larger $\tau_\mathrm{eq}$, similar to what was observed in Ref. \cite{Vovchenko:2016ijt}. However, the theoretical uncertainty due to the choice of linear or quadratic scaling for $\Gamma_{\text{inelastic}}$ is large enough that it is not possible to meaningfully quantify this effect. A more precise determination of how inelastic processes that produce thermal photons scale with the quark fugacity will be essential to use experimental data as a way to determine $\tau_\mathrm{eq}.$

\section{Summary and Outlook}

This work presented a model of chemical equilibration during the hydrodynamic phase of heavy-ion collisions that allows for a smooth transition from a pure glue initial state to a chemically equilibrated QGP. Integrated with a complete heavy-ion collision event generator, this allows one to simulate events where the abundance of (anti)quarks equilibrates according to an arbitrary timescale $\tau_\mathrm{eq}$. Our model assumes that particlization of the hydrodynamic fluid occurs at a quark fugacity-dependent transition temperature between 158 MeV and 260 MeV. We have shown that tuning $\tau_\mathrm{eq}$ has noticeable effects on hadronic observables, particularly on the development of flow, which is suppressed out of equilibrium. Charged particle multiplicities are surprisingly insensitive to chemical equilibration due to competing effects of the increased particlization temperature and reduced quark fugacity, although there is evidence for a limited degree of baryon suppression for long values of $\tau{eq}$. Thermal photon production is also sensitive to quark chemical equilibration, but better constraints on the fugacity dependence of photon production rates will be essential to allow for a meaningful model-to-data comparison.

We plan to extend this model further by studying differential flavor equilibration. In this paper, the construction of the EoS was simplified using the assumption that all quark flavors equilibrate simultaneously. Strangeness has long been of interest as a signature of the QGP \cite{Rafelski:1982pu}, so we will break this assumption and model strange quarks with a distinct fugacity and equilibration time from light quarks. This should allow for a more accurate study of strange hadrons such as the $\Omega$ baryon and $\phi$ meson, which are of particular interest as tests of QGP equilibration due to their very limited scattering after hadronization \cite{Dumitru:1999sf}.

We also propose to further study the effects of varying QGP system size. As shown in Sec. \ref{sec:evolution}, smaller collision systems such as O+O will likely equilibrate less than Pb+Pb before hadronization. There is additionally evidence in the present work from the effect on $v_2$ that less central events are more strongly impacted by introducing quark fugacities. By exploring a wider range of collision systems and centralities, we expect to see that the observable effects of a QGP hadronizing before fully equilibrating will depend on the QGP lifetime. This will provide a range of systems that should demonstrate greater sensitivity to the chemical equilibration timescale than seen in the Pb+Pb results of this work.

A primary aim of these future studies will be to use experimental data to constrain the equilibration times of quark flavors. The results presented here serve as a proof of concept that observables measured at the LHC and RHIC can, in principle, be used to distinguish between different equilibration timescales $\tau_\mathrm{eq}$. With additional observables that are sensitive to the flavor equilibration timescales, we can set the stage for a Bayesian analysis that rigorously constrains the timescales alongside other free parameters such as the QGP transport coefficients.

\section*{Acknowledgements}

 The authors thank Pierre Moreau for his invaluable help in constructing the equation of state in partial chemical equilibrium. We also thank Horst Stöcker for useful discussions. This work was supported by the U.S. Department of Energy Grants No. DE-FG02-05ER41367 (A.G., S.A.B., and B.M.) and No. DE-SC-0024347 (J.-F.P.). This research used resources of the National Energy Research Scientific Computing Center (NERSC), a U.S. Department of Energy Office of Science User Facility under Contract No. DE-AC02-05CH11231 us-
ing NERSC Award No. NP-ERCAP0026721.

\appendix
\label{appendix}

\section{Entropy Density}
\label{sec:s}

Taking the fundamental thermodynamic relation

\begin{align}
    dU = T dS - P dV + \mu dN
\end{align}

and considering only the total number of (anti)quarks $N_q$, we have

\begin{align}
    \varepsilon = T s - P + \mu_q n_q,
\end{align}

where $\mu_q$ is the (anti)quark chemical potential. Note that in chemical equilibrium, $\mu_q = 0$ and $s = (\varepsilon + P)/T$ follows immediately. Defining the quark fugacity as $\gamma_q = e^{\mu_q/T}$, Maxwell relations give

\begin{align}
    n_q = \left(\frac{\partial N}{\partial V}\right)_{\mu_q, T} = \left(\frac{\partial P}{\partial \mu_q}\right)_{V,T} = \left(\frac{\partial P}{\partial \gamma_q}\right)_{V,T}\left(\frac{\partial \gamma_q}{\partial \mu_q}\right)_{V,T} = \frac{\gamma_q }{T}\left(\frac{\partial P}{\partial \gamma_q}\right)_{V,T},
\end{align} 

and thus

\begin{align}
    s = \frac{\varepsilon + P}{T} - \frac{\gamma_q  \ln{\gamma_q}}{T} \left(\frac{\partial P}{\partial \gamma_q}\right)_{V,T}.
    \label{eq:spart}
\end{align}

Using Eq. \ref{eq:P}:

\begin{widetext}
\begin{align}
\left(\frac{\partial P}{\partial \gamma_q}\right)_{V,T}
= {}&
\left(\frac{T_\mathrm{c}}{T_3}\right)^4
P_3\!\left(T \frac{T_3}{T_\mathrm{c}}\right)
- \left(\frac{T_\mathrm{c}}{T_0}\right)^4
P_0\!\left(T \frac{T_0}{T_\mathrm{c}}\right)
\label{eq:dPdgamma} \\[0.5ex]
&+ \gamma_q
\left(\frac{4 T_\mathrm{c}^3}{T_3^4}\right)
\left(\frac{\partial T_\mathrm{c}}{\partial \gamma_q}\right)
P_3\!\left(T \frac{T_3}{T_\mathrm{c}}\right)
+ (1-\gamma_q)
\left(\frac{4 T_\mathrm{c}^3}{T_0^4}\right)
\left(\frac{\partial T_\mathrm{c}}{\partial \gamma_q}\right)
P_0\!\left(T \frac{T_0}{T_\mathrm{c}}\right) \notag \\
&+ \gamma_q
\left(\frac{T_\mathrm{c}}{T_3}\right)^4
P_3'\!\left(T \frac{T_3}{T_\mathrm{c}}\right)
\left(-\frac{T T_3}{T_\mathrm{c}^2}\right)
\left(\frac{\partial T_\mathrm{c}}{\partial \gamma_q}\right)
+ (1-\gamma_q)
\left(\frac{T_\mathrm{c}}{T_0}\right)^4
P_0'\!\left(T \frac{T_0}{T_\mathrm{c}}\right)
\left(-\frac{T T_0}{T_\mathrm{c}^2}\right)
\left(\frac{\partial T_\mathrm{c}}{\partial \gamma_q}\right).
\notag
\end{align}
\end{widetext}

Inserting this result into Eq. \ref{eq:spart} yields

\begin{widetext}
\begin{align}
s(T,\gamma_q) = {}&
\frac{\varepsilon(T,\gamma_q) + P(T,\gamma_q)}{T}
\label{eq:s} \\[0.5ex]
&- \frac{\gamma_q \ln \gamma_q}{T}
\left(
\left(\frac{T_\mathrm{c}(\gamma_q)}{T_3}\right)^4
P_3\!\left(T \frac{T_3}{T_\mathrm{c}(\gamma_q)}\right)
- \left(\frac{T_\mathrm{c}(\gamma_q)}{T_0}\right)^4
P_0\!\left(T \frac{T_0}{T_\mathrm{c}(\gamma_q)}\right)
\right) \notag \\
&- \frac{2 \sqrt{\gamma_q} \ln \gamma_q}{T}
\frac{T_3 - T_0}{T_\mathrm{c}(\gamma_q)}
\left(
\gamma_q
\left(\frac{T_\mathrm{c}(\gamma_q)}{T_3}\right)^4
P_3\!\left(T \frac{T_3}{T_\mathrm{c}(\gamma_q)}\right)
+ (1-\gamma_q)
\left(\frac{T_\mathrm{c}(\gamma_q)}{T_0}\right)^4
P_0\!\left(T \frac{T_0}{T_\mathrm{c}(\gamma_q)}\right)
\right) \notag \\
&+ \frac{\sqrt{\gamma_q} \ln \gamma_q}{2}
\frac{T_3 - T_0}{T_\mathrm{c}(\gamma_q)}
\left(
\gamma_q
\left(\frac{T_\mathrm{c}(\gamma_q)}{T_3}\right)^3
P_3'\!\left(T \frac{T_3}{T_\mathrm{c}(\gamma_q)}\right)
+ (1-\gamma_q)
\left(\frac{T_\mathrm{c}(\gamma_q)}{T_0}\right)^3
P_0'\!\left(T \frac{T_0}{T_\mathrm{c}(\gamma_q)}\right)
\right).
\notag
\end{align}
\end{widetext}

This result is then computed using the previously defined functions $\varepsilon(T, \gamma_q)$, $P(T, \gamma_q)$, and $T_\mathrm{c}(\gamma_q)$, together with the lattice-derived pressure functions $P_3(T)$ and $P_0(T)$.


%

\end{document}